\documentstyle[11pt,aaspp4]{article}
\def\cm{{\rm cm}}
\def\etal{{\rm et~al. }}
\def\hmpc{\;h^{-1}{\rm Mpc}}

\def\kms{{\rm \;km\;s^{-1}}}

\def\lya{Ly$\alpha$}
\def\btauh{\overline{\tau}_{\rm HI}}
\def\btauhe{\overline{\tau}_{\rm HeII}}
\def\btau{\overline{\tau}}
\def\tauh{\tau_{\rm HI}}
\def\tauhe{\tau_{\rm HeII}}
\def\fh{C_{\rm HI}}
\def\fhe{C_{\rm HeII}}

\begin{document}

\title{Intergalactic Helium Absorption in Cold Dark Matter Models}
\author{Rupert A.C. Croft$^{1,6}$, David H. Weinberg$^{1,7}$,
Neal Katz$^{2,3,8}$, and Lars Hernquist$^{4,5,9}$}

\footnotetext[1]{Department of Astronomy, The Ohio State University,
Columbus, OH 43210}
\footnotetext[2]{ Department of Astronomy, University of Washington, 
Seattle, WA, 98195}
\footnotetext[3]{ Department of Physics and Astronomy, 
University of Massachusetts, Amherst, MA, 98195}
\footnotetext[4]{Lick Observatory, University of California, Santa Cruz, 
CA 95064}
\footnotetext[5]{Presidential Faculty Fellow}
\footnotetext[6]{racc@astronomy.ohio-state.edu}
\footnotetext[7]{dhw@astronomy.ohio-state.edu}
\footnotetext[8]{nsk@kestrel.phast.umass.edu}
\footnotetext[9]{lars@helios.ucolick.org}
 
\begin{abstract}
Observations from the Hopkins Ultraviolet Telescope 
and the Hubble Space Telescope  have recently detected HeII absorption along 
the lines of sight to two high redshift quasars.  We use cosmological 
simulations with gas dynamics to investigate HeII absorption in the cold dark
matter (CDM) theory of structure formation.  We consider two $\Omega=1$ CDM 
models with different normalizations and one open universe ($\Omega_0=0.4$) 
CDM model. The simulations incorporate the photoionizing UV background 
spectrum computed by Haardt \& Madau (1996), which is based on the output of
observed quasars and reprocessing by the \lya\ forest.  The simulated gas 
distribution, combined with the Haardt \& Madau spectral shape, accounts for 
the relative observed values of $\btauh$ and $\btauhe$, the effective mean 
optical depths for HI and HeII absorption.  If the background intensity is as
high as Haardt \& Madau predict, then matching the absolute observed values of 
$\btauh$ and $\btauhe$ requires a baryon abundance larger (by factors 
between 1.5 and 3 for the various CDM models)
than our assumed value of $\Omega_b h^2=0.0125$.  The simulations
reproduce the evolution of $\btauhe$ over the observed redshift range,
$2.2 \lesssim z \lesssim 3.3$, if the HeII photoionization rate remains roughly
constant.  

HeII absorption in the CDM simulations is produced by a diffuse, fluctuating,
intergalactic medium, which also gives rise to the HI \lya\ forest.  Much of
the HeII opacity arises in underdense regions where the HI optical depth is
very low.  We compute statistical properties of the HeII and HI absorption
that can be used to test the CDM models and distinguish them from an 
alternative scenario in which the HeII absorption is caused by discrete,
compact clouds.  The CDM scenario predicts that a substantial amount of 
baryonic material resides in underdense regions at high redshift.  HeII 
absorption is the only sensitive observational probe of such extremely diffuse,
intergalactic gas, so it can provide a vital test of this fundamental 
prediction.
\end{abstract}
 

\section{Introduction}
The Lyman-$\alpha$ forest seen in quasar spectra (Lynds 1971;
Sargent et al.\ 1980) is produced by
absorption from diffuse hydrogen at high redshift. The big bang model predicts
that approximately
25\% of primordial baryonic matter should also be in the form
of helium, and  that there should therefore be corresponding absorption
at shorter wavelengths
from the \lya\ transition of singly ionized helium (HeII).
It has recently become possible to detect redshifted HeII absorption
with space-based, ultraviolet (UV) observations 
(Jakobsen et al.\ 1994; Davidsen, Kriss, \& Zheng 1996; Hogan,
Anderson, \& Rugers 1996).  
Recent hydrodynamic cosmological simulations of cold dark matter (CDM) models
have been remarkably successful in 
reproducing the observed properties of the HI \lya\ forest
(Cen et al.\ 1994; Zhang, Anninos, \& Norman 1995; 
Hernquist et al.\ 1996, hereafter HKWM; Miralda-Escud\'e et al. 1996;
Dav\'e et al.\ 1997; Zhang et al.\ 1997, Rauch et al.\ 1997). 
In this paper we examine whether simulations like those of HKWM can
also account for the observed
HeII absorption, and we consider their implications 
for the physical state of the absorbing gas.

The basic physics of intergalactic HeII absorption is elegantly
described by Miralda-Escud\'e (1993).  
The ambient background of UV radiation
produced by quasars and (perhaps) young galaxies keeps
the diffuse hydrogen of the \lya\ forest in a highly
photoionized state, and it also ensures that most diffuse helium is
singly or doubly ionized. However, the UV background intensity is
lower at the  ionization energy of HeII (4 Rydbergs, 228$\AA$) than it is
at the HI ionization energy (1 Rydberg, 912$\AA$), 
and HeII absorption can therefore be significant even
in regions where the HI density is very low.
HeII absorption may be the only practical tool for directly
observing regions that lie significantly below the cosmic mean
density, revealing gas whose hydrogen \lya\ absorption might be buried
in noise or removed in the process of continuum fitting. According to
gravitational instability models of structure formation, the voids
between galaxies should harbor a substantial portion of the baryonic matter in
the universe.  A homogeneous intergalactic medium (IGM) would produce
a uniform absorption trough in quasar spectra (Gunn \& Peterson 1965). 
In a realistic gravitational instability model, the matter in underdense
regions should instead produce a fluctuating continuum of absorption
(Reisenegger \& Miralda-Escud\'e 1995).

Once the hydrogen in the universe has been reionized, the neutral
helium fraction is expected to be small except in high density,
collapsed regions.  At fixed density and temperature, the 
fraction of HeII in highly photoionized gas is inversely proportional to the
photoionization rate,
\begin{equation}
\Gamma_{\rm HeII}=\int^{\infty}_{\nu_{\rm HeII}} d\nu 
\frac{4\pi J(\nu)}{h\nu}\sigma_{\rm HeII}(\nu),
\label{eqn:ghe}
\end{equation}
where $J(\nu)$ is the specific intensity of the background
at frequency $\nu$,
$\sigma_{\rm HeII}$ is the ionization cross section for HeII, and
$h\nu_{\rm HeII}=$ 4 Rydbergs.
The HI fraction is inversely proportional to 
\begin{equation}
\Gamma_{\rm HI}=\int^{\infty}_{\nu_{\rm HI}} d\nu 
\frac{4\pi J(\nu)}{h\nu}\sigma_{\rm HI}(\nu),
\label{eqn:gh}
\end{equation}
which is dominated by photons with energy between 1 and 4 Rydbergs.
Measurements of the mean HeII and HI \lya\ opacities can 
constrain the spectral shape of the UV background through
the ratio of these integrals, $\Gamma_{\rm HI}/\Gamma_{\rm HeII}$,
provided one has a model that specifies the density and temperature
structure of the absorbing gas.
The evolution of the UV background's intensity and shape 
can be tracked by the evolution of the mean opacities with
redshift.  The simulations provide a realistic model for the IGM
with the physical detail needed to exploit this approach, albeit
a model whose properties depend (as they should) on the parameters
of the underlying cosmological scenario.

For a given cosmological model, or indeed for any model IGM in which
most of the absorption arises in highly photoionized gas, the predicted
mean opacity depends on the 
parameter combination $\Omega_{b}^{2}/\Gamma$, where $\Omega_b$
is the baryon density parameter 
(e.g., Miralda-Escud\'{e} and Ostriker 1992). 
This scaling assumes that gas temperatures (and hence recombination 
rates) are unaffected by changes in $\Omega_b$ and $\Gamma$, an
assumption that we will revisit in \S 3.2.
The simulations allow us to ask whether a cosmological model is
consistent with the observed mean opacities given constraints on
$\Omega_b$ from big bang nucleosynthesis (e.g., Walker et al.\ 1991)
and on $\Gamma$ from
the observed quasar population (e.g., Haardt \& Madau 1996)
or the proximity effect (e.g., Bajtlik, Duncan, \& Ostriker 1988).
The simulations also predict distribution functions for HI and HeII opacities
and for the ratio of these opacities, which can be compared to
observations in order to test the simulated IGM.

The possiblity of detecting HeII absorption from diffuse intergalactic gas
was one of the main scientific motivations for the Hopkins Ultraviolet
Telescope (HUT), which first flew on the Astro-1 mission in 1990
(Davidsen, private communication). While the quasar absorption experiment
could not be carried out during Astro-1, because of pointing problems, it
was successfully performed during the Astro-2 mission in 1995, as described by
Davidsen, Kriss \& Zheng (1996, hereafter DKZ). 
The first detection of HeII absorption was in fact achieved before Astro-2 by 
Jakobsen et al.\ (1994, hereafter JBDGJP),
who used the HST Faint Object Camera to observe the quasar Q0302-003 (z=3.28).
They measured a clear drop in the received quasar continuum across the
304$\AA$ (rest-frame) edge, and they inferred a high HeII optical depth,
$\tau_{\rm HeII}> 1.7$ at 90\% confidence.
Because of the relatively low spectral resolution,
the authors were unable to establish whether the absorption was
caused mainly
by material associated with individually identified HI lines or by a more
diffuse component. JBDGJP,
Giroux, Fardal, \& Shull (1995), 
Madau \& Meiksin (1995),
and Songaila, Hu, \& Cowie (1995)
have explored the implications of these data for the shape of the 
UV background spectrum assuming different analytic models of the 
absorbing medium.  If the absorption is dominated by discrete lines,
then the spectrum must be quite soft, but a harder spectrum is allowed
if most HeII absorption arises in a diffuse, ``Gunn-Peterson'' background.
We qualify this latter term with scare-quotes because the analytic
models typically assume a uniform Gunn-Peterson effect,
while the cosmological simulations predict a smoothly fluctuating, diffuse IGM
that blurs the traditional distinction between Gunn-Peterson
absorption and the \lya\ forest (HKWM; 
see also Miralda-Escud\'e \& Rees 1993; Cen et al.\ 1994; 
Miralda-Escud\'e et al. 1996).

Q0302-003 was recently reobserved by Hogan et al.\ (1996, hereafter HAR), 
using the Goddard High Resolution Spectrograph on HST.
HAR confirm JBDGJP's detection of HeII
absorption, but the higher spectral resolution of the GHRS observations
reveals interesting new details.
The transmission below the HeII break remains fairly high ($\sim 0.3$)
within about 4000$\kms$ of the quasar.  HAR attribute the
relatively low HeII fraction in this region to photoionization caused
by Q0302-003 itself, with the ionization zone terminated by a high
column density HI absorber.  They find a low but significant level
of residual flux at shorter wavelengths
corresponding to $\tauhe=2.0^{+1.0}_{-0.5}$ at 95\% confidence.
The upper limit on $\tauhe$ implies that helium remains predominantly
doubly ionized even outside of the ionization zone produced by Q0302-003, 
presumably because of the ambient UV background.
The upper limit depends on accurate background subtraction at the
blue end of the spectrum, a challenging problem that is 
discussed in detail by HAR.  

HST cannot probe HeII absorption below $z\approx 3$.
DKZ took advantage 
of the shorter wavelength sensitivity of the 
HUT to measure HeII absorption in the spectrum of HS1700+64 (z=2.743). 
They find a mean opacity in the redshift range $2.2 < z < 2.6$
corresponding to $\tau_{\rm HeII}=1.0\pm0.07$, 
with some evidence that the opacity increases as a function of redshift
over this range.
The observed wavelength interval is $\sim 150\AA$ and the spectral resolution
$\sim 3\AA$, so the HUT spectrum contains enough information to
reveal structure in the residual flux.  (The spectrum shown in DKZ
is averaged in $10\AA$ bins.)  
The analysis in this paper will be aimed primarily at the
HUT observations.
The physical issues are rather different for Q0302-003 because of
the important role of the observed quasar itself in ionizing the
absorbing gas.  We will therefore save a detailed comparison to the 
JBDGJP and HAR observations for
a future paper.

The numerical approach in this paper will be similar to that of HKWM,
who use smoothed-particle hydrodynamics (SPH) simulations to study
the HI \lya\ forest in a critical density, CDM universe.
Studies of HeII (and HI) absorption in Eulerian-grid hydrodynamic
simulations have been carried out by Zhang et al.\ (1995, 1996) and
Miralda-Escud\'e et al.\ (1996).
Here we will study the physical state of the gas that produces HeII absorption
in CDM-dominated, gravitational instability models of structure formation,
relating it to and differentiating it from the gas that dominates
HI absorption. We will use information
on HeII and HI in the context of these models to study the required UV
background spectrum, its evolution with redshift, and the implied baryon
density. We will examine several variants of the CDM scenario,
enabling us to  see which features
are generic within this cosmological picture and which can be used as
diagnostics  for constraining cosmological parameters.
We will also  compare the simulation results to those of  a
simple model where all of the flux is absorbed in discrete lines.  
This sort of model is often used as a phenomenological description of the
HI \lya\ forest, though it does not correspond physically to
what happens in the simulations, where the absorbing structures are
relatively diffuse and merge continuously into a fluctuating background.
We obtain predictions that can be compared to 
future observational analyses that probe
HI and HeII absorption along a common line of sight. 

\section{Simulations}
We have used 
the N-body plus SPH code TreeSPH 
(Hernquist \& Katz 1989; Katz, Weinberg, \& Hernquist 1996, hereafter KWH)
to simulate three different CDM-dominated cosmological models,
the parameters of which are listed in columns~$2-5$ of Table~1.
The first is a ``standard'' CDM (SCDM) universe, 
with $\Omega=1$,
$h=0.5$ (where $h \equiv H_{0}/100\;\kms\;{\rm Mpc}^{-1}$), and
$\Omega_{b}=0.05$. The power spectrum is normalized so that the rms 
amplitude of mass fluctuations in $8\hmpc$ spheres, linearly
extrapolated to $z=0$, is $\sigma_{8}=0.7$.
This normalization is consistent with
that advocated by White, Efstathiou \& Frenk (1993)
to match the observed masses of rich galaxy clusters,
but it is inconsistent with the normalization implied by the
COBE-DMR experiment.
Our second model is identical to the first
except that $\sigma_{8}=1.2$. This 
higher amplitude is consistent with the 4-year
COBE data (Bennett et al.\ 1996), and we therefore label the model CCDM.
The third model, OCDM, assumes an open universe with 
$\Omega_{0}=0.4$, $h=0.65$ and $\Omega_{b}=0.03$.
The transfer functions used are those of Efstathiou, Bond \& White 1992.
The shape parameter $\Gamma=0.234$ was used for the OCDM model,
which is also COBE-normalized ($\sigma_{8}=0.75$, Gorski et al.\ 1995).
The baryon fraction for these models, $\Omega_{b}=0.0125 h^{-2}$, was
chosen based on the big bang nucleosynthesis analysis of 
Walker et al.\ (1991), who deduce primordial abundances of
D, $^3$He, $^4$He, and $^7$Li from local observations 
using models for chemical processing of elements in stars. 
Measurement of the deuterium abundance in high redshift Lyman limit
systems offers a more direct route to determining $\Omega_b h^2$.
The first applications of this method have so far given results
which favor $\Omega_{b}h^2$ being a factor 
of $\sim 2$ smaller (Carswell et al.\ 1994; Songaila et al.\ 1994;
Rugers \& Hogan 1996ab, but see also Hogan 1997)
 or a factor of $\sim 2$ larger 
(Tytler, Fan \& Burles  1996, Tytler, Burles \& Kirkman 1996, but see also
Songaila, Wampler \& Cowie 1996).
The observational situation is thus uncertain at present,
though we shall see that 
the high $\Omega_{b}h^2$ results are favored in the CDM models.

\begin{table}[b]
\centering
\caption[junk]{\label{models} Parameters of cosmological models. The
conventions are defined in \S 2, except for $C$ which is defined in equation
(7).}
\vspace{0.5cm}
\begin{tabular}{cccccccc}
model & $\Omega_0$ & $h$ & $\Omega_b$ & $\sigma_8$ & 
$\fhe$ $(z=2-3.3)$ &$ \fh (z=2)$ & $\fh (z=3)$ \\ 
\hline &&&\\
SCDM & 1 & 0.5 & 0.05 & 0.7 & 3.0 & 2.6 & 2.8\\
CCDM & 1 & 0.5 & 0.05 & 1.2 & 8.0 & 5.7 & 6.7\\
OCDM & 0.4 & 0.65 & 0.03 & 0.75 & 3.0 & 2.4 & 2.8\\
\end{tabular}
\end{table}

A periodic cubic volume of comoving side length $11.111 \hmpc$ was simulated
for each model, each simulation having identical random phases and being
evolved to $z=2$. We analyzed
outputs at $z=3.5, 3.0, 2.67, 2.33$ and $2$, encompassing the range of
existing HeII absorption observations.
Each simulation was run with $64^{3}$ collisionless
dark matter particles and $64^{3}$ gas particles.
A spatially uniform photoionizing radiation field was imposed, and
radiative cooling and heating rates were calculated assuming photoionization 
equilibrium and optically thin gas, as discussed in KWH.
The spectral shape of the background radiation field 
and its evolution with redshift were taken from the work of
Haardt \& Madau (1996, hereafter HM).
This background was calculated by HM
from a self-consistent treatment of the absorption and re-emission
of light from observed quasars by the \lya\ forest.

The implementation of photoionization in TreeSPH requires values of
the photoionization rates and photoionization heating rates for
HI, HeI, and HeII as a function of redshift 
(see Katz, Weinberg, \& Hernquist 1996).  We compute these parameters
from the HM spectrum at intervals $\Delta z=0.05$, and for the
simulations we use fitting formulae that match these values to
within $\sim 10\%$.  The parameters that are particularly relevant 
for this paper are the photoionization rates $\Gamma_{\rm HI}$
and $\Gamma_{\rm HeII}$, for which we adopt
\begin{equation}
 _{\rm HM}\Gamma_{\rm HI} = 
 1.115 \times 10^{-12} \exp\left[-0.57565(z-2.5)^2\right]\,{\rm s}^{-1},
\label{fit_gh}
\end{equation}
\begin{equation}
 _{\rm HM}\Gamma_{\rm HeII} = 
 1.088 \times 10^{-14} \exp\left[-0.57565(z-2.5)^2\right] P(z)\,{\rm s}^{-1},
\label{fit_ghe}
\end{equation}
where the factor
\begin{equation}
 P(z) = 1 + 0.125(z-2.5)+0.0825(z-2.5)^2
\end{equation}
accounts for the slight difference in the relative evolution of
$\Gamma_{\rm HI}$ and $\Gamma_{\rm HeII}$.
At $z \sim 2-4$,
the values implied by equations~(\ref{fit_gh}) and~(\ref{fit_ghe})
are about 30\% lower than those shown in figure~6 of HM
because we fit HM's $q_0=0.5$ results rather than their
$q_0=0.1$ results (see \S 5.13 of HM).
We set the UV background to zero at $z>6$, since observations suggest
that the population of quasars is already declining rapidly between
$z=4$ and $z=5$.  
As shown by Hui \& Gnedin (1997), the thermal state of the IGM
at $z \la 3$ is insensitive to the epoch of reionization
provided that it occurs at $z \ga 5$.
If HI or HeII reionization occurs at $z<5$, the IGM could be somewhat
hotter than our simulations predict, a point that we will return
to at the end of \S 3.4.

The HKWM results suggested that a simulation with the HM background
and $\Omega_b=0.05$ would underproduce the observed HI \lya\ opacity,
and we therefore reduced the amplitude of the HM background by 
a factor of two (i.e. to half the values implied by equations 3 and 4)
before evolving the simulations.
We can recover the impact of different UV background intensities
at the analysis stage, regardless of the specific value adopted 
during dynamical evolution, as discussed below in \S 3.1.
The SCDM simulation of this paper is identical to the simulation
analyzed by HKWM except for the UV background spectrum
(HKWM used a $\nu^{-1}$ power-law) and the inclusion of star
formation (using the prescription of KWH). 
Star formation and the associated feedback only influence the
simulation results in high density regions that are not
the focus of this paper.

The UV background determines the relative fractions of different
ionic species at specified density and temperature.
Figure~1 shows the fractions of HI, HeI, and HeII as a function
of temperature for (total) hydrogen densities 
$n_H=3.95 \times 10^{-6}\;\cm^{-3}$ (left panel) and
$3.95 \times 10^{-5}\;\cm^{-3}$ (right panel).
These are, respectively, the mean density and an overdensity of 10
at $z=2.33$ for $\Omega_b h^2=0.0125$.
We will focus much of our analysis on this redshift because it is
the simulation output closest to the middle of DKZ's observed redshift range.
Abundances are computed by requiring balance between the production
and destruction rates for each species, as described in KWH.
Thin lines show abundances for no UV background --- these
collisional equilibrium abundances depend on temperature alone
and are thus the same in the two panels.
Thick lines show abundances for the HM background, with intensity
reduced by a factor of three (relative to HM), the value required
for our SCDM and OCDM models to reproduce the observed HeII opacity
(see \S 3.1).  Photoionization dominates completely over collisional
ionization at low temperatures, where the HI and HeII fractions fall
as $T^{-0.7}$ because of the temperature dependence of the recombination rates.
Collisional ionization becomes significant at high temperatures.
Raising the density increases the recombination rates and, therefore,
the HI, HeI, and HeII fractions, though these never exceed their collisional
equilibrium values.  Lowering the UV background by a constant factor
would have the same effect on these plots as raising the density
by the same factor.  

The {\it shape} of the background
spectrum determines the relative fractions of HI and HeII
through the ratio of photoionization rates, 
$\Gamma_{\rm HI}/\Gamma_{\rm HeII}$.
HM assume a $\nu^{-1.5}$ power-law for the intrinsic UV spectrum of
their quasar sources, but the ambient background that they 
compute is substantially softer because of reprocessing by the \lya\ forest.
In particular, the spectrum at $z \sim 2.5$ drops by a factor $\sim 15$
at $h\nu \sim 4$~Rydbergs because of HeII absorption.
If we adopted a pure $\nu^{-1.5}$ power-law with intensity chosen
to produce the HI fractions in Figure~1, the HeII fractions at
low temperatures would be reduced by a factor of $3.3$.
With the $\nu^{-1}$ power-law of HKWM, the reduction would be a 
factor of $6.5$.  It is worth noting, however, that these
changes in equilibrium abundances have no dynamical impact on the
simulations and minimal impact on the ability of the gas to cool
in collapsed objects.  The fraction of gas that cools and condenses
into galaxies is nearly identical in simulations with the HM
background and a $\nu^{-1}$ background (Weinberg, Hernquist, \& Katz 1996).

\begin{figure}
\vspace{9.0cm}
\includegraphics{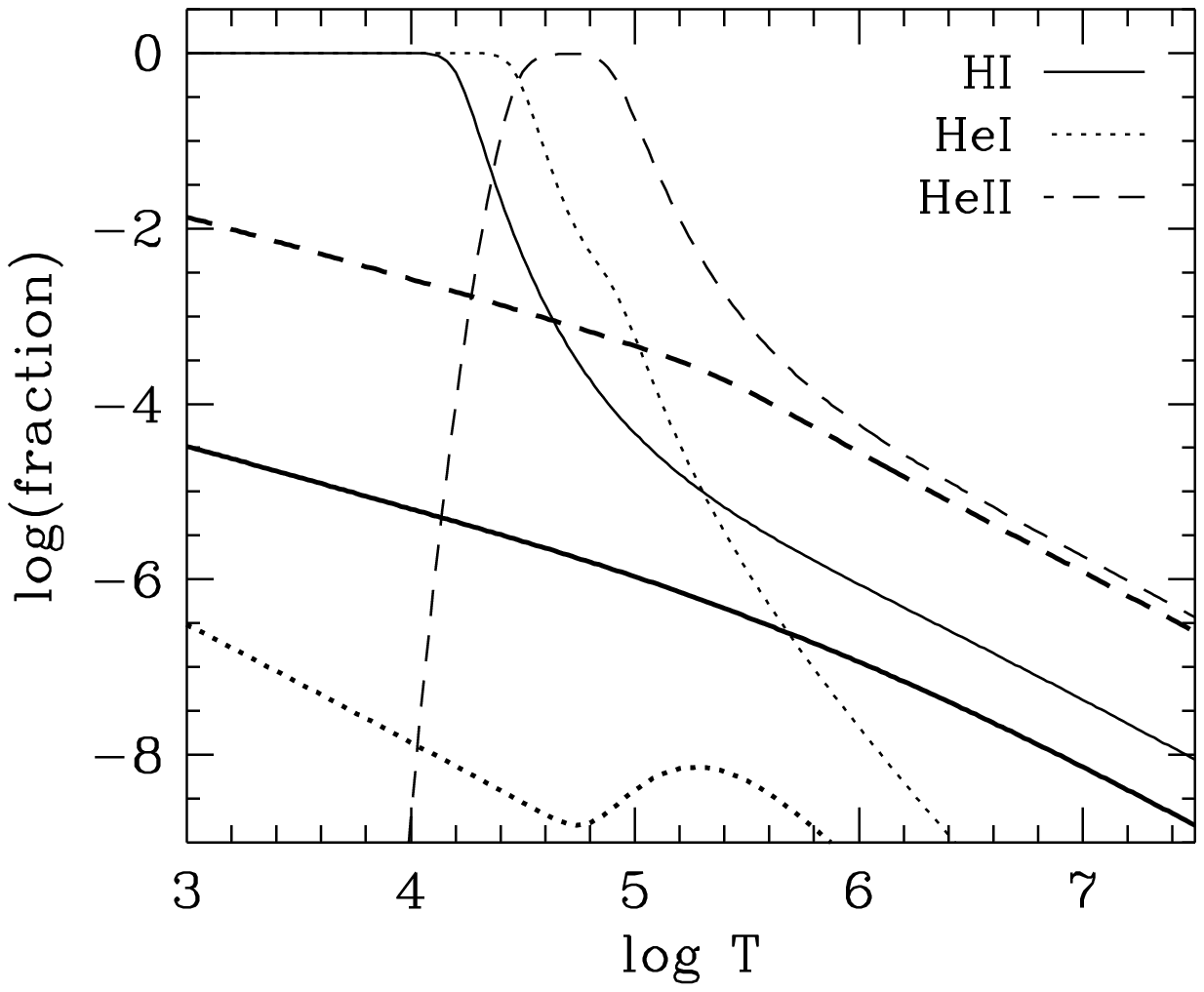}
\includegraphics{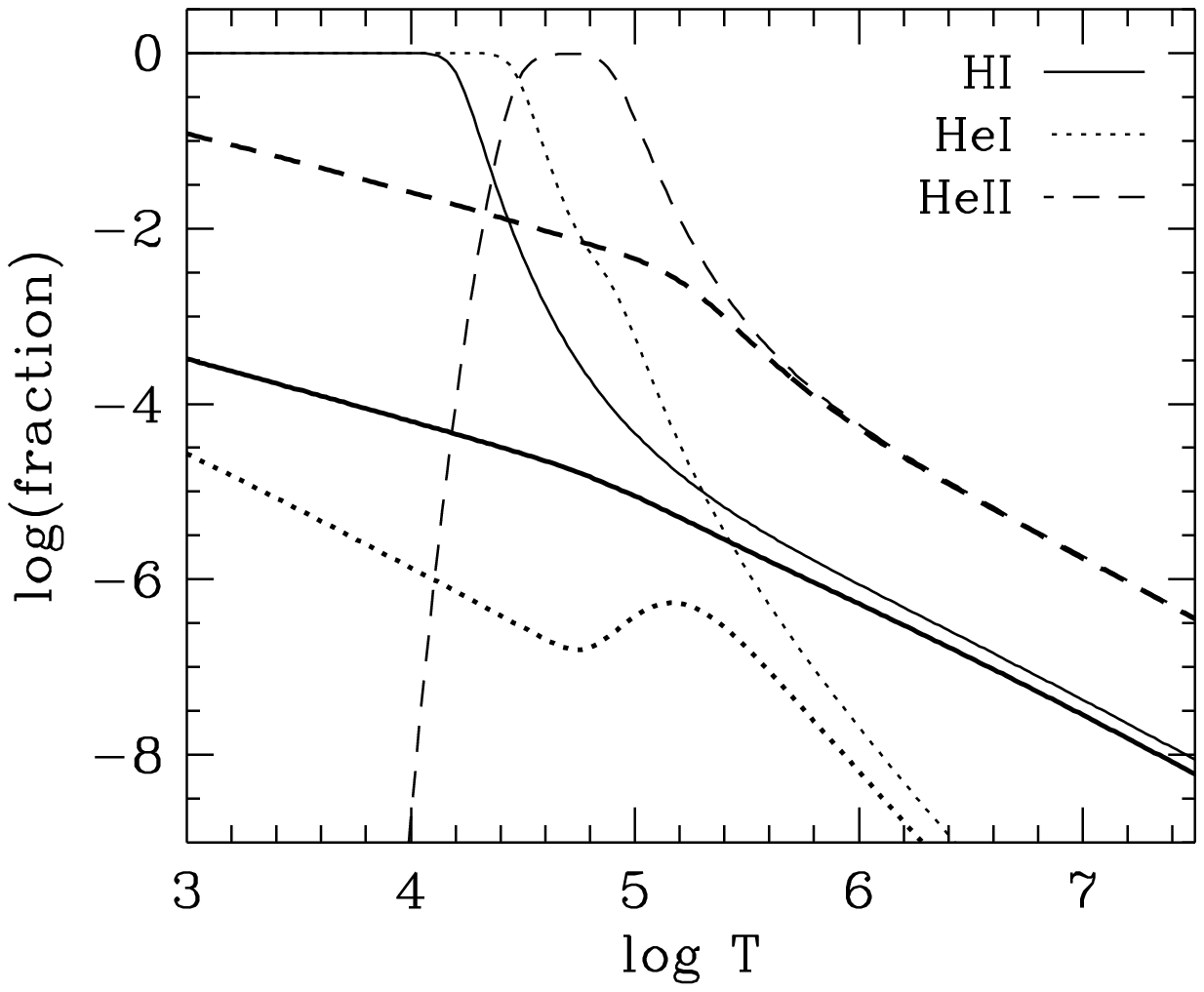}
\caption[junk]{\label{ionfrac}
Fractions of HI (solid lines), HeI (dotted lines), and HeII (dashed lines)
as a function of gas temperature.
The assumed gas densities are $n_H=3.95 \times 10^{-6}\;\cm^{-3}$ 
in the left panel (the mean density at $z=2.33$ for $\Omega_b h^2=0.0125$)
and a factor of ten higher in the right panel.
Thin lines show collisional equilibrium abundances.  Thick lines
show abundances in photoionization equilibrium with a UV background
that has the spectral shape computed by HM but an intensity
lower by factor of three.
}
\end{figure}

\section{Results}
The \lya\ optical depth produced by a uniform medium with neutral 
hydrogen or singly ionized helium density $n$ is
(Gunn \& Peterson 1965; Miralda-Escud\'e 1993)
\begin{equation}
\tau = \frac{\pi e^2}{m_e c} f\lambda H^{-1}(z) n,
\label{eqn:gp}
\end{equation}
where $H(z)$ is the Hubble constant at redshift $z$,
$f=0.416$ is the \lya\ oscillator strength,
and $\lambda$ is the transition wavelength
($1216\AA$ and $304\AA$ for HI and HeII, respectively).
In the simulations, the intergalactic gas is not uniform, so the
optical depth varies as a function of wavelength and angular position.
Figure~2 shows HI and HeII ``absorption maps'' of a slice through
the center of the SCDM simulation at $z=2.33$.
(the velocity width of the full cube is $2029\;\kms$ at this redshift).
The faintest grey levels correspond to $\tau \sim 0.05$, a value 
commonly adopted as a threshold for line identification in high 
signal-to-noise spectra.  Absorption saturates for $\tau \ga 3$
($e^{-\tau} \la 0.05$).

\begin{figure}
\vspace{9.0cm}
\includegraphics{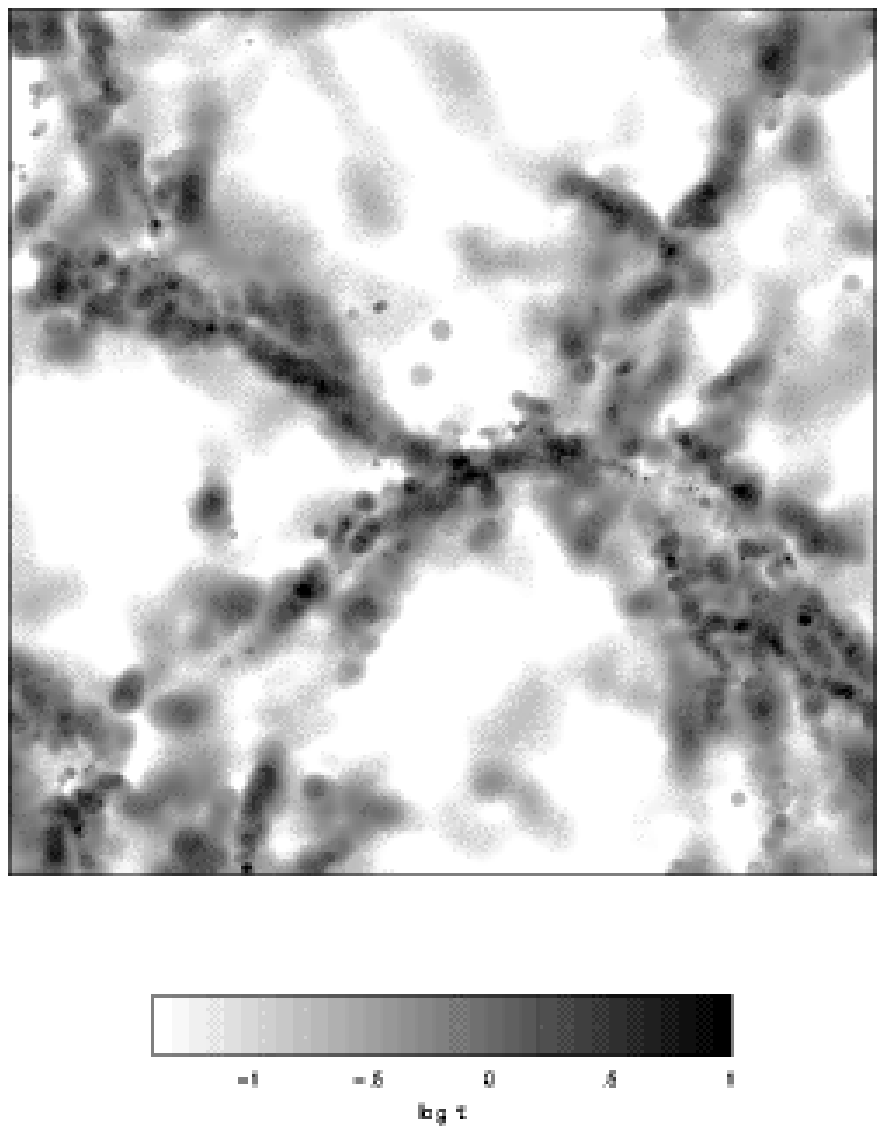}
\includegraphics{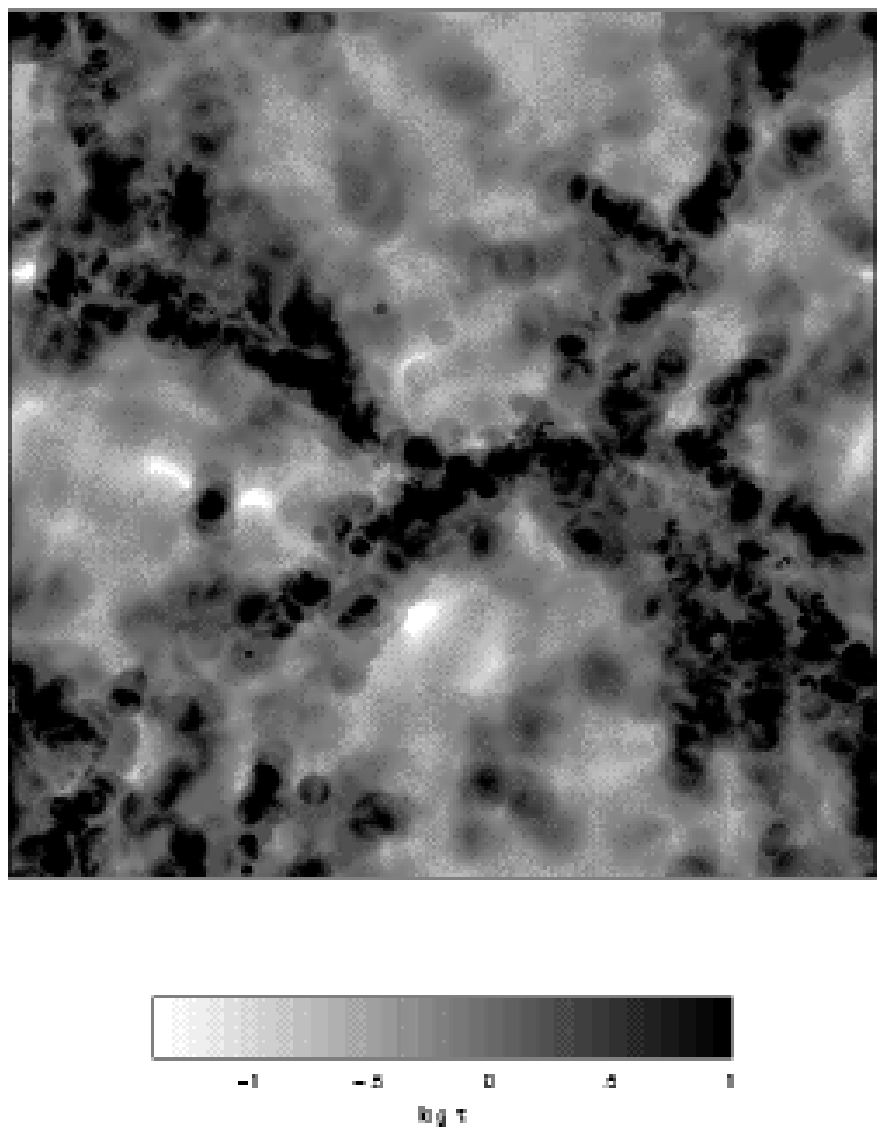}
\caption[junk]{\label{heproj}
Optical depth to HI (left) and HeII (right) absorption in the
SCDM simulation at $z=2.33$.  The optical depth is averaged
over a slice of velocity width 60 $\kms$ through the middle of the 
simulation cube.  The scale is logarithmic in $\tau$, as indicated,
with the faintest grey levels corresponding to $\tau \sim 0.05$.
The intensity of the UV background has been scaled so that the 
simulation produces the mean HeII opacity measured by DKZ (see discussion
in \S 3.1).
HeII absorption is prominent in low density regions that are nearly
transparent in HI.
}
\end{figure}

As expected, these maps show stronger HeII absorption
than HI absorption.  HeII produces a substantial optical depth
in regions that are nearly devoid of HI absorption,
so HeII measurements can probe low density structure that is 
virtually undetectable by other means.
Conversely, most regions with a significant HI optical depth produce
saturated HeII absorption.  The sharp edges visible in some of the
saturated HeII features and the ``blobbiness'' of the weak HeII
absorption are probably artifacts of representing the
gas distribution with a discrete set of particles.

For quantitative analyses, we create
simulated absorption spectra along random lines of sight through 
the simulation cube.
Knowing the temperature and density of each SPH particle,
we can compute the associated fractions of HI, HeI, and HeII
given our assumed UV background.  From these we compute the
optical depths as a function of frequency by performing a line
integral through the smoothing kernels of all SPH particles
whose kernels intersect the line of sight, taking into account
Hubble flow, peculiar velocities, and the thermal broadening
appropriate to each of the absorbing species
in order to transform from physical space to frequency space.
Details of this procedure are described in HKWM.

\subsection{Mean absorption and evolution with redshift}
Figure~3a shows the redshift evolution of the effective mean
optical depth for HeII absorption, 
$\btauhe \equiv -\log_{e} \langle F \rangle$, where 
$F$ is the transmitted flux (with $F=1$ corresponding to complete
transmission).  
The average is performed over 100 lines of sight at each of the five
output times. 
Because of the non-linear dependence of flux
on optical depth, $\btauhe$
is not the mean value of $\tauhe$ along the spectrum,
but while $\langle F \rangle$ can be measured from a spectrum
of imperfect resolution and finite signal-to-noise ratio, 
the mean value of $\tauhe$ cannot.
Thin lines show results for the three models with the UV
background employed during dynamical evolution of the simulation,
i.e., the HM background divided by a factor of two.
The intensity of this UV background at energies 
responsible for ionizing HeII is approximately constant 
over the redshift range plotted here, though it
does start to decline smoothly  above $z\sim 3.5$. 
The mean optical depth decreases towards lower redshift primarily because 
the expansion of the universe reduces the HeII fraction, since it
lowers the physical density of the absorbing medium and hence the
recombination rate per HeIII ion.  Cosmic expansion also
spreads the atoms in a given comoving region over 
a larger range in frequency. 
These are precisely the effects that determine the 
evolution of the Gunn-Peterson optical depth produced by a uniform
IGM with a constant UV background.
The first effect, which dominates here,
would not apply to gas in collapsed, physically
stable structures. While Figure 2 shows that the absorption is by no means
uniform, the rapid evolution of $\btauhe$ is a strong hint that the 
absorbing gas resides in low density structures that are still
expanding with the cosmic background.

\begin{figure}
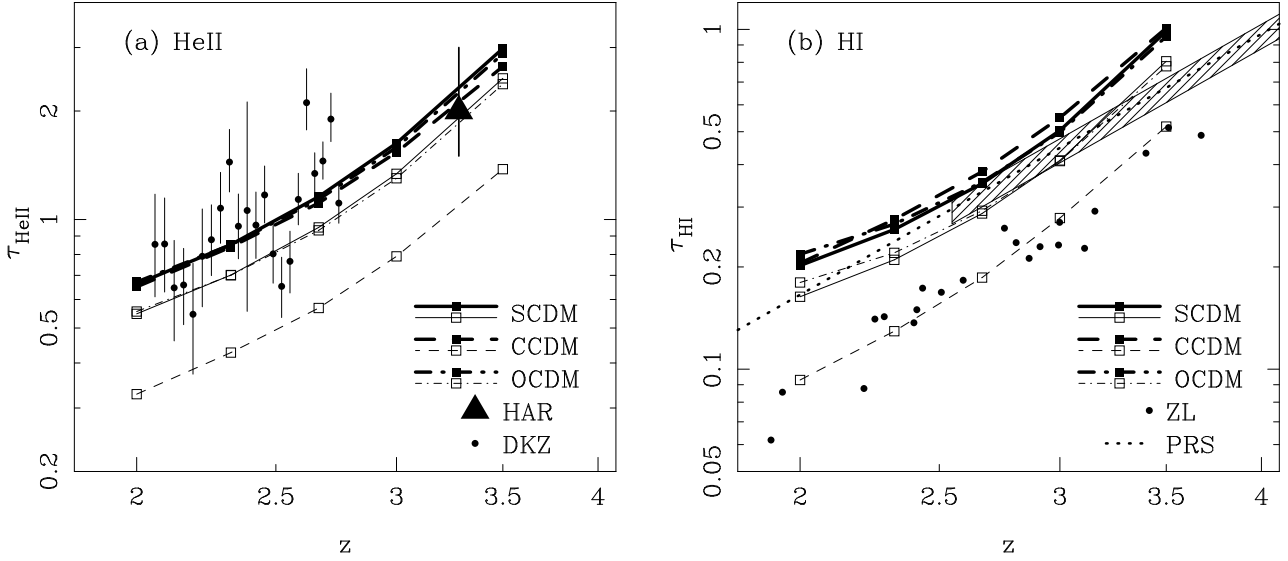

\vspace{9.0cm}
\includegraphics{f3a.ps}
\includegraphics{f3b.ps}
\caption[junk]{\label{he2vsz}
{\it (a)} Redshift evolution of $\btauhe$, the effective mean optical
depth of HeII absorption.  The horizontal axis (redshift) is scaled
logarithmically in $(1+z)$, so that power-law relations between $\tau$
and $(1+z)$ would appear as straight lines.  
Thin lines with open squares show the results for the
three CDM simulations calculated with the ionizing
background intensity used during dynamical evolution.
Thick lines with filled squares show $\btauhe$ after
the intensity of the background is reduced by a factor of 1.5 for SCDM
and OCDM and by 4.0 for CCDM.  With this 
lower intensity, the simulations reproduce the mean optical depth measured
by DKZ, $\btauhe=1.0$ at $z=2.4$.  Points from the
individual $10\AA$ bins of the DKZ spectrum
are shown as solid dots with 1$\sigma$ error bars.
The triangle shows the measurement and $2\sigma$ error bar of HAR,
$\btauhe=2.0^{+1.0}_{-0.5}$ at $z \approx 3.3$.
{\it (b)} Same as {\it (a)}, but for HI absorption.
Thin lines again correspond to the UV background used during dynamical
evolution and thick lines to the background reduced by a factor of
1.5 (SCDM, OCDM) or 4.0 (CCDM).
The dotted line is PRS's fit to the observed
evolution of $\btauh$, which was calculated from data in the redshift
range indicated by the hatched region. The vertical extent of this region 
illustrates the approximate $1\sigma$ confidence interval 
for the fit. The solid points are the measurements of mean optical depth
from ZL for their ``sample 2'' quasars. 
}
\end{figure}

The solid triangle in Figure~3a shows HAR's observational estimate,
$\btauhe=2.0^{+1.0}_{-0.5}$ (2$\sigma$ errors) at $z=3.3$,
measured outside the ionization zone produced by the observed quasar.
The wavelength range contributing to this point is about $40\AA$.
Solid circles with error bars show $\btauhe$ at lower redshifts from
the DKZ spectrum.  Each point corresponds to a single $10\AA$ bin
from DKZ's figure~1 --- we measured the extrapolated quasar
continuum and observed flux from this figure and divided them to 
obtain $F={\rm exp}(-\btauhe)$.  The error bars are $1\sigma$ and are
based on the noise vector in DKZ's figure.  The fluctuations from 
point to point are larger and more coherent than expected from 
observational errors alone.
They are presumably caused by large scale
density variations, or possibly  by
inhomogeneity in the background radiation field.

Comparison between the data points and the thin lines in Figure 3a shows
that the CDM models produce roughly the correct trend of
$\btauhe$ with redshift but have an overall level of absorption that is
too low.  The intensity of the ionizing background used
during the dynamical evolution of the simulations is 
already a factor of two lower than that
advocated by HM, but a  further reduction is
necessary to match the observed mean absorption. 
To quantitatively compare the models, we calculate the factor
\begin{equation}
\fhe=\frac {_{\rm HM}\Gamma_{\rm HeII}} {_{\rm req}\Gamma_{\rm HeII}},
\end{equation}
where $_{\rm req}\Gamma_{\rm HeII}$ is the photoionization rate required to 
match the observed HeII optical depth and $_{\rm HM}\Gamma_{\rm HeII}$
is the photoionization rate predicted by the HM background (equations 3 and 4).
Changing the intensity of the UV background within a broad range
does not significantly alter the temperatures or spatial distribution of the
absorbing gas, so we are able to estimate 
$_{\rm req}\Gamma_{\rm HeII}$ by rescaling the background
after the simulation has been run but before creating simulated spectra
(see Miralda-Escud\'e et al., in preparation, for numerical tests
of this scaling).  In fact, because
photoionization completely dominates over collisional ionization
in the absorbing regions, we could obtain virtually identical
results by the still simpler procedure of rescaling optical depths in the
extracted spectra.

The HeII absorption value we  match is the quoted value of 
$\btauhe=1.0\pm0.07$ at $z=2.4$ from DKZ. We find $\fhe$
by an iterative interpolation search, dividing the HM
ionization parameters by different factors until we
obtain the correct mean absorption. As we do not have a simulation output at 
$z=2.4$, we linearly interpolate $\btauhe$
between $z=2.33$ and $z=2.67$.
The values of $\fhe$ for the three cosmological models are listed in column~6
of Table~1.  These factors are equal for the two low
amplitude models, OCDM and SCDM. 
The required value for the CCDM model is much higher.
This model tends to produce less
absorption at fixed $\Gamma$ because (as we will show in detail later)
most of the flux decrement arises in large volumes 
that are near or below the mean density.
At the higher mass fluctuation amplitude of CCDM,
more of the gas has flowed out of these
void-like regions into higher density zones and collapsed objects;
the absorption from these regions is already saturated, so adding
more gas to them does not increase the mean decrement.

The heavy lines in Figure 3a show $\btauhe$ with the rescaled ionization
parameters. A single value of $\fhe$ allows
the models to fit the observed  $\btauhe$ results at all redshifts from 
$z=2$ to $z=3.3$, to within the scatter of the data points,
indicating that the evolution predicted by the simulations is
consistent with the observations if the UV background evolves as
predicted by HM.  Although $\Gamma_{\rm HeII}$ falls by only 20\%
between $z=2.4$ and $z=3.3$, the models reproduce the factor of two
difference between the DKZ and HAR measurements of $\btauhe$
because of the cosmic expansion effects discussed earlier.

Figure 3b shows the effective mean optical depth of HI absorption,
$\btauh \equiv -{\rm log}_e\langle F_{\rm HI} \rangle$,
for the three CDM models.  Thin lines correspond to the UV background
intensity adopted in the simulations, i.e., half of the HM background
intensity at the corresponding redshift.  Thick lines show $\btauh$
with the UV background intensity rescaled in order to match the observed
$\btauhe$, i.e., the HM background (equations 3 and 4)
divided at all redshifts by the factor
$\fhe$ listed in Table 1.

The straight dotted line in Figure 3b shows the power-law fit
$\btauh = 0.0037(1+z)^{3.46}$ found by Press, Rybicki, \& Schneider
(1993, hereafter PRS) in their analysis of the mean flux decrement
in a sample of 29 high-redshift quasars.  The hatched box shows the
redshift extent of points used to compute this fit and the $1\sigma$
statistical uncertainty in the fit.  PRS computed the mean decrement
by extrapolating the quasar continuum from the region redward of
\lya\ emission into the \lya\ forest region.  Zuo \& Lu (1993, hereafter ZL)
estimated the mean decrement from higher resolution spectra by directly
fitting the continuum to the regions of lowest absorption in 
the forest.  We indicate their results by the filled circles in 
Figure 3b.  Clearly the two observational determinations disagree
by far more than their statistical uncertainties.
A recent analysis of Keck HIRES spectra, using an approach similar
to ZL's but data of higher resolution and signal-to-noise
ratio, yields mean decrements that are much closer
to the PRS values than to ZL's (Rauch et al.\ 1997).
We will therefore proceed on the assumption that the PRS determination
is accurate, but the systematic uncertainty in existing estimates
of $\btauh$ is worth keeping in mind, as a major change to these
estimates would affect our conclusions about the shape of the UV
background spectrum.

Once the background intensity has been divided by the factor
$\fhe$, the effective HI optical depths agree fairly well with
the PRS determination, though they tend to rise above it at the low and 
high ends of our redshift range.  To better quantify this agreement,
we list in columns 7 and 8 of Table 1 the scaling factors $\fh$
(defined analogously to $\fhe$) by which the HM background must
be divided in order that the simulation match
the PRS optical depths at $z=2$ and $z=3$.  
These factors match the corresponding values
of $\fhe$ to 20\% or better in most cases (30\% for CCDM at $z=2$),
indicating that the cosmological simulations and the HM spectral shape
are, taken together, consistent with the joint observational
constraints of DKZ and PRS.   The HI scaling factors at $z=2$ and
$z=3$ are also similar, indicating that the simulations reproduce the
PRS evolution law over this redshift range if the UV background
evolves as predicted by HM.  If this analysis is extended to $z \sim 4$,
however, the simulations require a roughly constant $\Gamma_{\rm HI}$
(HKWM; Rauch et al.\ 1997),
while the HM model predicts a substantial drop in $\Gamma_{\rm HI}$
towards high redshift because of the declining number density of
quasar sources.

The uncertainty on the correct value of $\fhe$ is mainly statistical,
and is dominated by the fact that an observational
 measurement of $\btauhe$ is only  available from one QSO.  The theoretical
estimates at z=2.33 come from 200 lines of sight through the simulation box,
which together cover a redshift patch
 equal to $\sim 13$ times the useful length of the DKZ spectrum.
We can estimate the error in  $\fhe$ by picking groups of simulated
spectra with the same total length as DKZ from our ensemble of 200. We then 
calculate the  $\fhe$ required to match the observational $\btauhe$
for each set of spectra. From the spread of $\fhe$ values we estimate the
$1 \sigma$ uncertainties on $\fhe$ to be $+10 \%, -25\%$. The error in $\fh$
is dominated by systematic uncertainties in the observational determination
of $\btauh$, probably associated with continuum fitting (see Rauch et al.
1997).

JBDGJP suggested that the high HeII optical depth
towards Q0302-003 might arise because the HeII ``Stromgren spheres''
around quasars had not overlapped by $z=3.3$, leaving most of the universe
optically thick to HeII ionizing photons.  Supporting evidence for this
scenario comes from a rapid change at $z \sim 3.1$ in
the SiIV/CIV ratios measured in \lya\ absorbers, which
suggests a change in the shape of the UV
background spectrum that could correspond to percolation of 
quasars' HeII ionization zones (Songaila \& Cowie 1996, but see also
Hellsten et al 1997). 
Our results show that, in the CDM models, no major change in the
spectral shape is needed to explain the existing HeII data.
Indeed, if HAR's upper limit on $\btauhe$ is taken at face value, then the
background spectrum cannot be much softer than the HM spectrum at $z=3.3$,
at least along this line of sight.
If HAR's residual flux is an artifact of imperfect background subtraction,
then our models could also accommodate HeII reionization at $z<3.3$.

\subsection{Implications for $\Omega_b$}
All of our models require $\fh \ga 2.5$ to match the PRS
determination of $\btauh$ and $\fhe \ga 3$ to match the DKZ
determination of $\btauhe$.  If we had obtained 
values of $\fh$ and $\fhe$ smaller than unity from the simulations,
we could accommodate them easily by appealing to additional
UV sources not considered by HM, e.g., star-forming galaxies or faint
AGNs.  However, since the HM background is based on the observed
population of quasar sources (with a modest extrapolation for quasars
below existing survey detection limits), it is difficult to see
how the true background could be lower than the HM background by
such large factors.  Reductions of this magnitude would also make
the background intensity inconsistent with estimates from the proximity
effect (Giallongo et al.\ 1996 and references therein),
though these are subject to significant systematic
uncertainties.  Even allowing for the imperfect resolution of the simulations
and plausible uncertainties in the background intensity, it seems that
these cosmological models for the \lya\ forest are at best
marginally compatible with the PRS opacity measurements if
$\Omega_b h^2=0.0125$.

The alternative to lowering the UV background is to raise the 
mean baryon density.  The \lya\ optical depth is proportional to
the number density of absorbing atoms (see equation~[\ref{eqn:gp}]),
which for highly photoionized gas is proportional to the
square of the gas density divided by the photoionization rate.
The second power of density arises because the
the recombination rates per HII or HeIII ion are themselves
proportional to the density.  
If the distribution of {\it over}densities ($\rho_b/\bar\rho_b$)
and gas temperatures in the IGM is unchanged by altering $\Omega_b$,
then the optical depth $\tau$ at a specified redshift along a line
of sight is proportional to $\Omega_b^2/\Gamma$.  Note that the
effective mean optical depth $\btau$ is not simply proportional
to $\Omega_b^2/\Gamma$ because of the non-linear nature of flux 
averaging, but it is still the case that raising $\Omega_b$ by
a factor $C^{1/2}$ should have the same effect as lowering $\Gamma$
by a factor $C$.  The assumption that the absorbing gas is
highly photoionized breaks down in regions that are collisionally
ionized (e.g., hot gas in virialized halos) or predominantly neutral
(e.g, damped \lya\ systems), but these are too rare to make much
contribution to $\btau$.

We have completed one simulation of the SCDM model with 
$\Omega_b=0.125$ instead of 0.05, and we find that the scaling
of $\tau$ with $\Omega_b$ is weaker than the above argument would
suggest, roughly $\tau \propto \Omega_b^{1.7}$ instead of 
$\tau \propto \Omega_b^2$.  
The reason is that raising $\Omega_b$
also raises the gas temperature in the low and moderate density
regions ($\rho_b/\bar\rho_b \la 10$) that produce most of the
absorption, because increasing the HI and HeII fractions allows
a given volume of gas to absorb energy from the photoionizing
background at a higher rate.  Since the recombination rates decline
as $T^{-0.7}$ in the relevant temperature regime, the HI and HeII
fractions do not rise by the full $\Omega_b^2$ factor when $\Omega_b$
is increased.  The physics of the $\Omega_b$ scaling does not
depend on the cosmological scenario, and we therefore expect the
result derived from our pair of SCDM simulations to hold more
generally (but see the discussion of reionization effects at the 
end of \S 3.4).
One might think that reducing $\Gamma$ at fixed
$\Omega_b$ would also alter the gas temperatures, but it does not,
because the increase in HI and HeII fractions is exactly compensated
by the smaller rate of photoionizations per ion.  
The scaling $\tau \propto \Gamma^{-1}$ is therefore preserved unless the
{\it shape} of the ionizing background, which determines the mean residual
energy per photoelectron, is altered.

With these scalings in mind, we can relate the values of $\fh$ and
$\fhe$ listed in Table 1 to the combination of $\Gamma$, $\Omega_b$,
and $h$ that is required for the simulation to match the observed
mean opacity:
\begin{equation}
C=\left(\frac{\Gamma_{\rm HM}}{\Gamma_{\rm true}}\right)
\left(\frac {\Omega_{b}h^{2}}{0.0125}\right)^{1.7}
\left(\frac{h_{\rm sim}}{h}\right),
\label{eqn:omegab}
\end{equation}
where $h_{\rm sim}$ is the value of $h$ adopted in the simulation.
The $h$ dependences arise because the mean gas density is proportional
to $h^2$ at fixed $\Omega_b$ and because the optical depth at fixed HI
or HeII density is inversely proportional to the Hubble constant
(equation~[\ref{eqn:gp}]).
If we assume that $\Gamma_{\rm true}=\Gamma_{\rm HM}$ and $h=h_{\rm sim}$,
then matching the PRS values of $\btauh$ requires
$\Omega_b h^2 \sim 0.023$ for SCDM and OCDM and $\Omega_b h^2 \sim 0.038$
for CCDM.  Matching the DKZ measurement of $\btauhe$ requires a
similar baryon density, though the HeII opacity on its own gives
a less compelling argument for high $\Omega_b$ because the lower limit
on $\Gamma_{\rm HeII}$ is less secure than the lower limit on 
$\Gamma_{\rm HI}$.
Assuming standard big bang nucleosynthesis, a density 
$\Omega_b h^2 = 0.023$ is in excellent agreement with 
Tytler et al.'s (1996) estimate
of the primordial deuterium abundance, but it is inconsistent with
the much higher deuterium abundances estimated by 
Carswell et al.\ (1994), Songaila et al.\ (1994), and
Rugers \& Hogan (1996ab).
We examine the baryon density required by cosmological simulations of
the \lya\ forest more thoroughly in Rauch et al. (1997), 
which includes a new determination of $\btauh (z)$ from high resolution
spectra and a discussion of the lower limit on $\Gamma_{\rm HI}$.
The analytic arguments presented in Weinberg et al.\ (1997) show that the 
lower limits on $\Omega_{b}$ derived from this method depend
 only on very general properties of the ``cosmological'' picture of
the Ly$\alpha$  forest, and are unlikely to be weakened substantially 
by changes in the adopted cosmological model or the numerical resolution
of the simulations.

\subsection{Simulated HeII spectra}
The top panels of Figure 4 show examples of HI (solid lines),
HeI (dotted lines), and HeII (dashed lines) absorption along two
randomly selected lines of sight through the SCDM simulation,
at $z=2.33$.  We have scaled the HM ionizing background
by the factor $\fhe$ listed in Table 1, so that the mean optical
depth matches the DKZ observation.  The transmission $e^{-\tau}$
is plotted against line-of-sight velocity.  The corresponding
baryon density (in units of the mean baryon density) is plotted below
each spectrum in the second panel, with the solid line showing the
redshift space density and the dotted line the real space density
(i.e., the density computed with peculiar velocities and thermal
broadening set to zero).  The HI optical depth is well correlated
with the redshift space density.  Features in the redshift space density
field are usually offset from those in the real space field because
of peculiar motions, and the high density peak in the left hand
spectrum is greatly broadened in redshift space because of infall.
HeI absorption is non-negligible
only in the highest density region of the second spectrum, where the
high recombination rate increases the relative fraction of neutral helium.
HeII absorption, on the other hand, is quite strong, and most of it arises
in regions where the HI optical depth is quite low.  
The ratio $\tau_{\rm HeII}/\tau_{\rm HI}$
(shown by the solid line in the third panel) is about a factor
of eight over most of the spectrum.
The HI and HeII flux decrements, $1-e^{-\tau}$, 
are nearly equal when the HI optical depth is high, but
when $\tau_{\rm HI}$ is small the HeII flux decrement is eight times
higher.  HeII absorption therefore probes regions
of lower density than HI absorption, as already seen in Figure~2.

\begin{figure}
\vspace{8.5cm}
\includegraphics{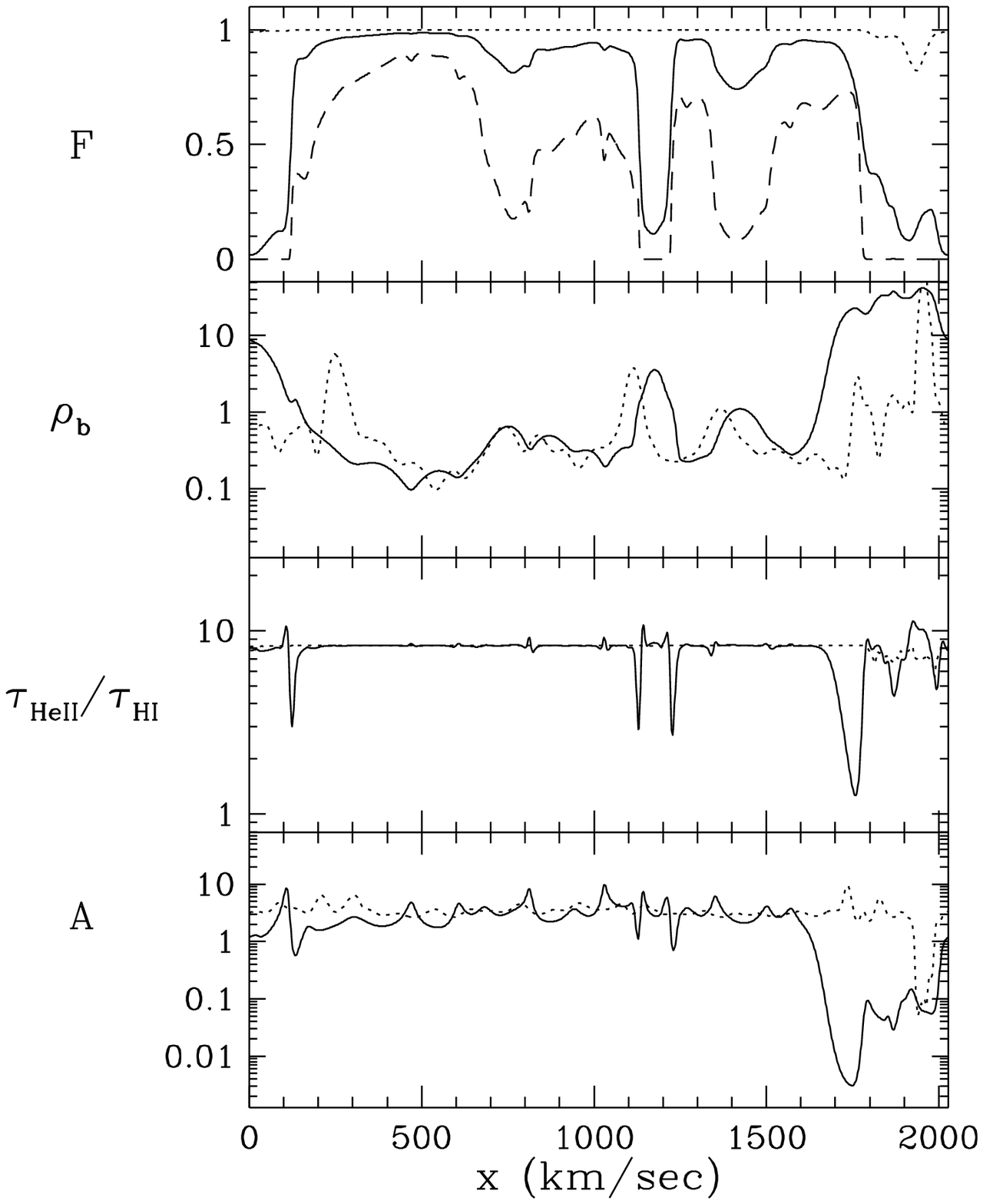}
\includegraphics{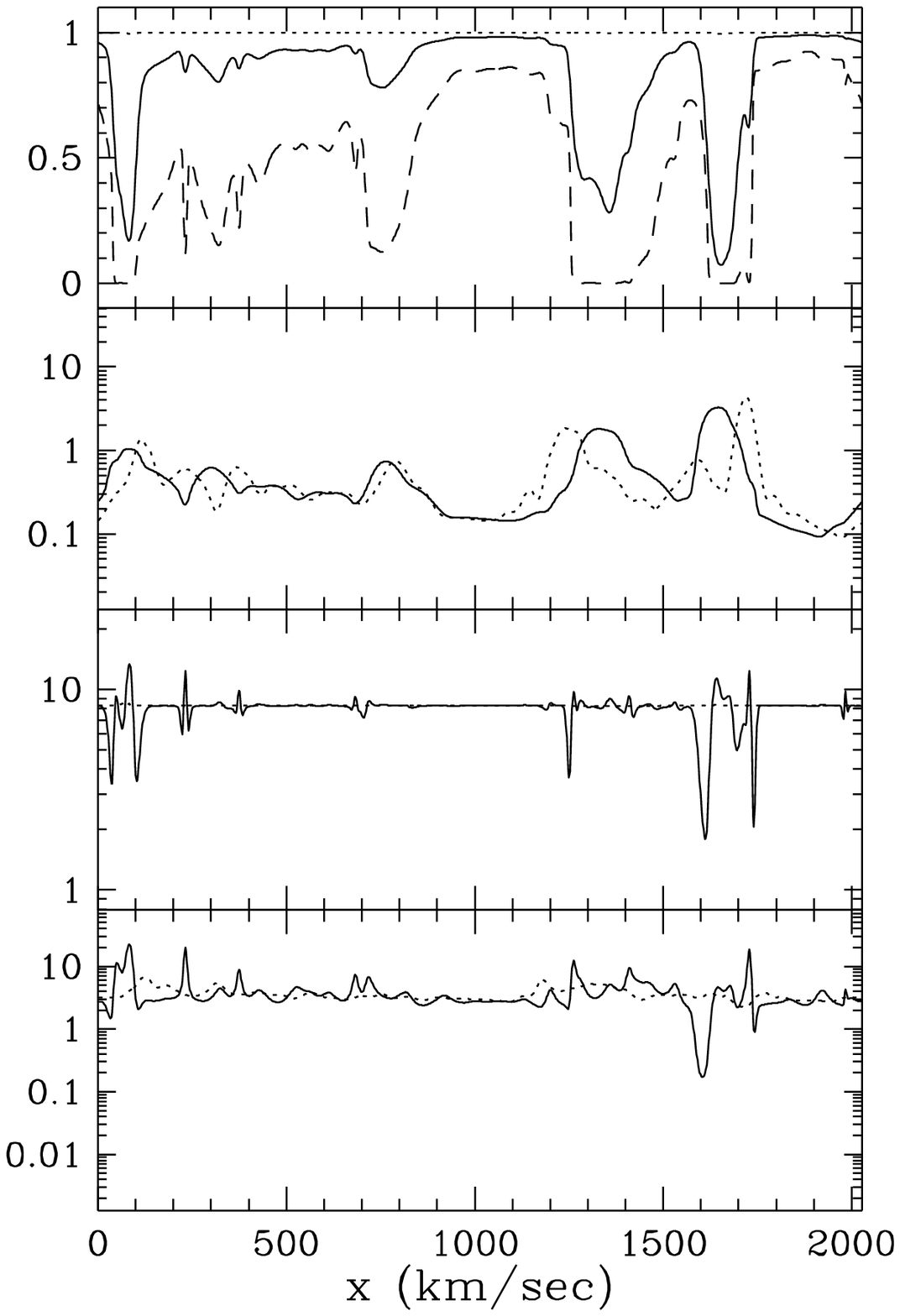}
\caption[junk]{\label{scdm0.7spec}
Two examples of spectra taken along random 
lines of sight through the simulation box
of the SCDM model, at a redshift $z=2.33$. The top panel
shows transmitted flux against line-of-sight 
velocity for HI (solid line), HeI (dotted line), and HeII (dashed line).
The second panel shows the density  
of baryonic matter, in units of the mean baryon density, in 
redshift space (solid line) and real space (dotted line).
The third panel shows the ratio $\tau_{\rm HeII}/\tau_{\rm HI}$
computed with (solid line) and without (dotted line) thermal broadening
of the absorption.  The bottom panel shows
the ratio $A=\tau_{\rm HeII}/\rho_{b}^{1.6}$ computed in 
redshift space (solid line) and real space (dotted line).
}
\end{figure}

\begin{figure}
\vspace{10.9cm}
\includegraphics{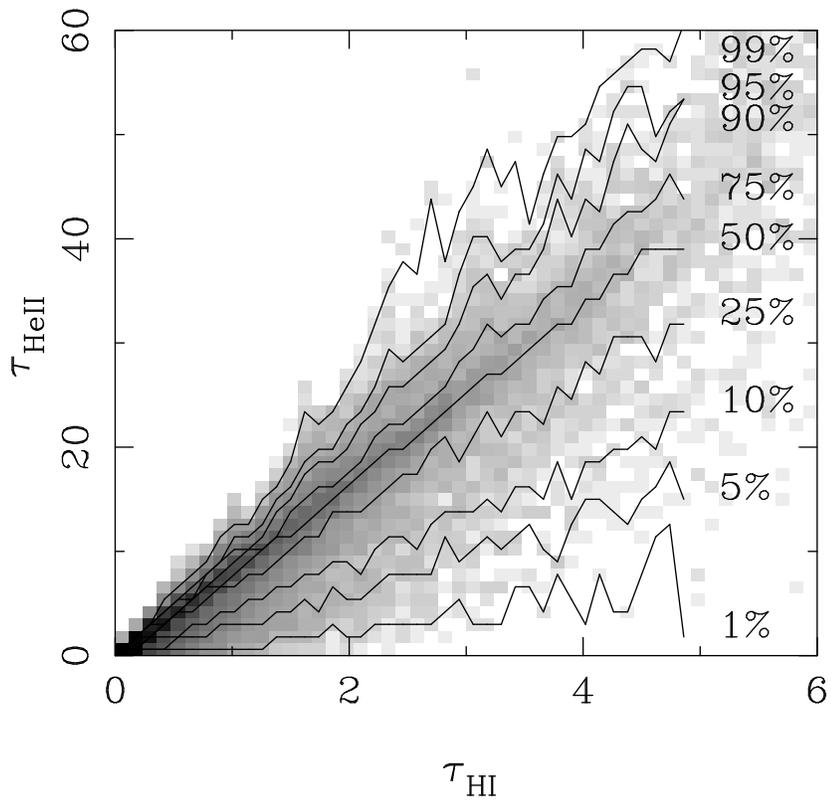}
\caption[junk]{\label{thresh}
The joint distribution of $\tauhe$ and $\tauh$ in the SCDM model at 
$z=2.33$, computed from 200 spectra, each with 1000 pixels.
The logarithmic grey scale shows the fraction of pixels in
bins of $\Delta \tauh=0.12$, $\Delta \tauhe =1.2$.
Lines show the percentile distribution of $\tauhe$ in each bin of $\tauh$.
}
\end{figure}

\begin{figure}
\vspace{10.9cm}
\includegraphics{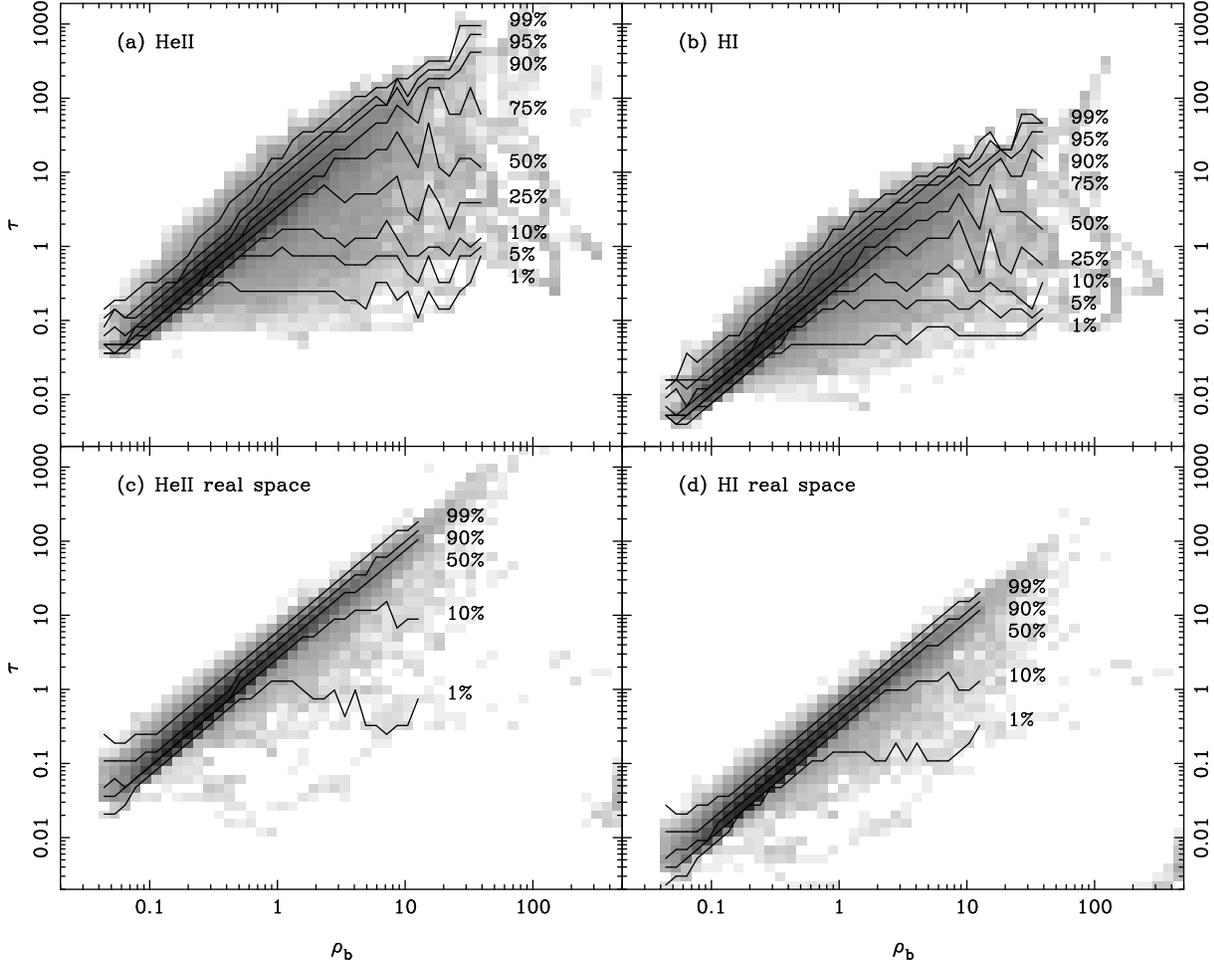}
\caption[junk]{\label{thres}
The joint distribution of optical depth and $\rho_b$ (in units of the mean
baryon density) for the SCDM model at $z=2.33$, 
in a format similar to Fig.~5.
{\it (a)} HeII in redshift space.  
{\it (b)} HI in redshift space.  
{\it (c)} HeII in real space.  
{\it (d)} HI in real space.  
}
\end{figure}

The optical depth ratio $\tauhe/\tauh$ is high and approximately 
constant, as anticipated by Miralda-Escud\'e (1993).  
Variations arise when collisional
ionization or thermal broadening of the spectrum become important ---
these processes affect HeII and HI differently because of the
differences in ionization potential and atomic mass, respectively.
The dotted line in the third panel of Figure 4 shows the optical depth
ratio calculated from spectra along the same lines of sight with
no thermal broadening applied.
The variations seen in the solid line largely disappear, indicating
that thermal broadening is their primary cause, at least along
these two lines of sight.  

For a more quantitative view of the optical depth ratio, we
examine the joint distribution of $\tauhe$ and $\tauh$ in 
200 spectra extracted along random lines of sight through the
SCDM simulation at $z=2.33$.  The logarithmic grey scale in Figure~5
indicates the distribution of pixels in the $\tauh - \tauhe$ plane.
There are $2 \times 10^5$ pixels in total, 1000 in each of the 
200 spectra.  The labeled lines in Figure~5 show the percentile
ranges of $\tauhe$ in bins of $\Delta \tauh=0.12$, 
i.e., of the pixels that have a given value of $\tauh$, 1\% have
$\tauhe$ below the 1\% line, 5\% below the 5\% line, and so forth.
Most pixels lie along a well defined ridge
at $\tauhe=8\tauh$, tracked by the median $\tauhe$ line. 
Collisional ionization in shock heated regions
raises $\tauhe /\tauh$ because it 
suppresses HI absorption more strongly than HeII absorption.
Thermal broadening on the edges of high density regions tends to reduce 
$\tauhe /\tauh$ by spreading HI absorption into the lower density surroundings.
Both effects can be seen in Figure~5, but the scatter below the median
relation is broader than the scatter above it, confirming the
anecdotal evidence of Figure~4 that thermal broadening is the
dominant cause of variations in $\tauhe/\tauh$.

Most of the absorbing gas in Figure 4 has relative density
$\rho/\bar\rho$ between 0.1 and 10.  In these simulations, the gas
in this density regime typically follows a tight and simple relation
between temperature and density, approximately $T \propto \rho^{0.6}$
(see figure~3b of Weinberg, Hernquist \& Katz 1997). 
This relation arises because
the gas is cooled by adiabatic cooling and heated by photoionization,
at a rate that depends on the density.  Changing the gas temperature
alters the heating rate, and gas that lies off this relation evolves
towards it on a Hubble timescale (for a more detailed 
discussion see Hui \& Gnedin 1996).
The relation breaks down in collapsed regions, where
shock heating and radiative cooling become important.
The optical depth to HI or HeII \lya\ absorption is proportional
to the number density of HI or HeII atoms at the corresponding
line-of-sight velocity.  These are proportional to the gas density 
multiplied by the recombination rate, which is in turn proportional
to $\rho T^{-0.7}$ for temperatures up to several$\;\times\;10^4\;$K.
Gas that lies on the $T \propto \rho^{0.6}$ temperature-density relation 
should therefore satisfy 
$\tauhe \propto \rho^2 (\rho^{0.6})^{-0.7} \propto \rho^{1.6}$.
The solid line in the bottom panel of Figure 4 plots the 
ratio $A=\tauhe /\rho^{1.6}$ against
velocity, and it is indeed approximately constant over most of the spectrum.
The variations are caused primarily by peculiar velocity distortions
of the redshift space density, since it is the real space density that
is directly correlated with the temperature.  In real space 
(dotted line) the variations are much smaller, with the one major
departure occurring in the high density region of the left hand
spectrum, where shock heating drives the gas off the simple 
temperature-density relation.
Figure~6a shows the joint distribution of $\tauhe$ and $\rho$,
in redshift space, for the 
200 spectra examined in Figure~5.  Most pixels lie on the ridge
$\tauhe \approx 3.5(\rho/\bar\rho)^{1.6}$, though there is a scatter towards
lower $\tauhe$ at higher densities because of shock heating.
The joint distribution of $\tauh$ and $\rho$ (Figure~6b) is similar,
except for a factor of eight offset.
The factor of two interquartile scatter at low densities in these plots is 
caused predominantly by peculiar velocity effects, as one can
see from the corresponding real space joint distributions (Figures~6c 
and~6d).

To a first approximation, one can thus regard a HeII or HI absorption
spectrum as a map of the gas density field along the line of sight,
albeit a map that is non-linear (an exponential of a power-law)
and distorted by peculiar velocities.
Figures 4--6 use the SCDM model for illustration, but the qualitative
physical picture is similar in all three of the cosmological scenarios
that we consider.  This picture can be contrasted with a traditional
phenomenological description of the \lya\ forest as a collection of
discrete absorbing clouds, each producing a Voigt-profile line 
fully characterized by a redshift, an HI column density, and a 
$b$-parameter (velocity width).  One can compute the HeII absorption
produced by ``line blanketing'' in such a model by assuming a
ratio $n_{\rm HeII}/n_{\rm HI}$ and either thermal broadening
(in which case the HeII $b$-parameters are a factor of two smaller than
the HI $b$-parameters)
or ``turbulent'' broadening (in which case the $b$-parameters are equal).
For a heavily saturated but undamped line, the
equivalent width is proportional to the $b$-parameter.

In order to compare the simulation results to this sort of discrete
cloud model, we have used a program written by J.  Miralda-Escud\'{e} 
to generate artificial spectra that are superpositions of randomly
distributed, Voigt-profile lines.
We draw HI column densities from a power-law distribution,
$dN/dN_{\rm HI} \propto N_{\rm HI}^{-1.5}$, with a lower cutoff at
$N_{\rm HI,min}=10^{12}\;\cm^{-2}$, 
and $b$-parameters from a Gaussian distribution with
a mean of $28\;\kms$ and a dispersion of $10\;\kms$, truncated
below $b_{\rm min}=18\;\kms$.
These parameters, based on Hu et al.\ (1995),
are similar to those used by Songaila et al.\ (1995)
in their modeling of the JBDGJP observation,
though we have lowered $N_{\rm HI,min}$ from $2\times 10^{12}\;\cm^{-2}$
to $10^{12}\;\cm^{-2}$.  We choose the mean number of lines per unit
redshift in order to match the PRS determination of $\btauh$ at $z=2.33$.
We then generate two sets of corresponding HeII spectra, one for pure
thermal broadening, one for pure turbulent broadening, choosing the
$n_{\rm HeII}/n_{\rm HI}$ ratio in each case to reproduce the
DKZ value of $\btauhe$ at $z=2.33$.  

\begin{figure}
\vspace{5.0cm}
\includegraphics{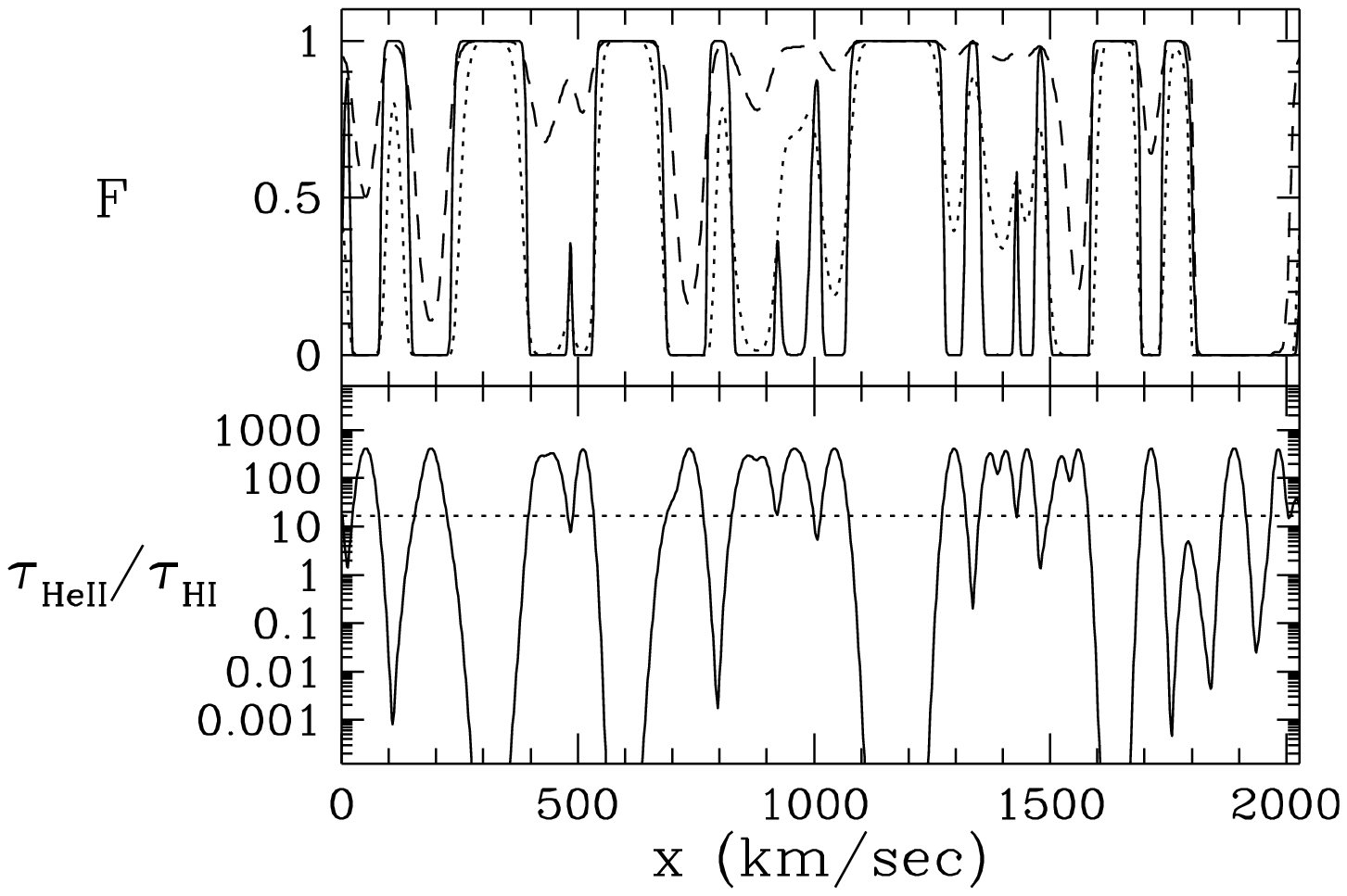}
\includegraphics{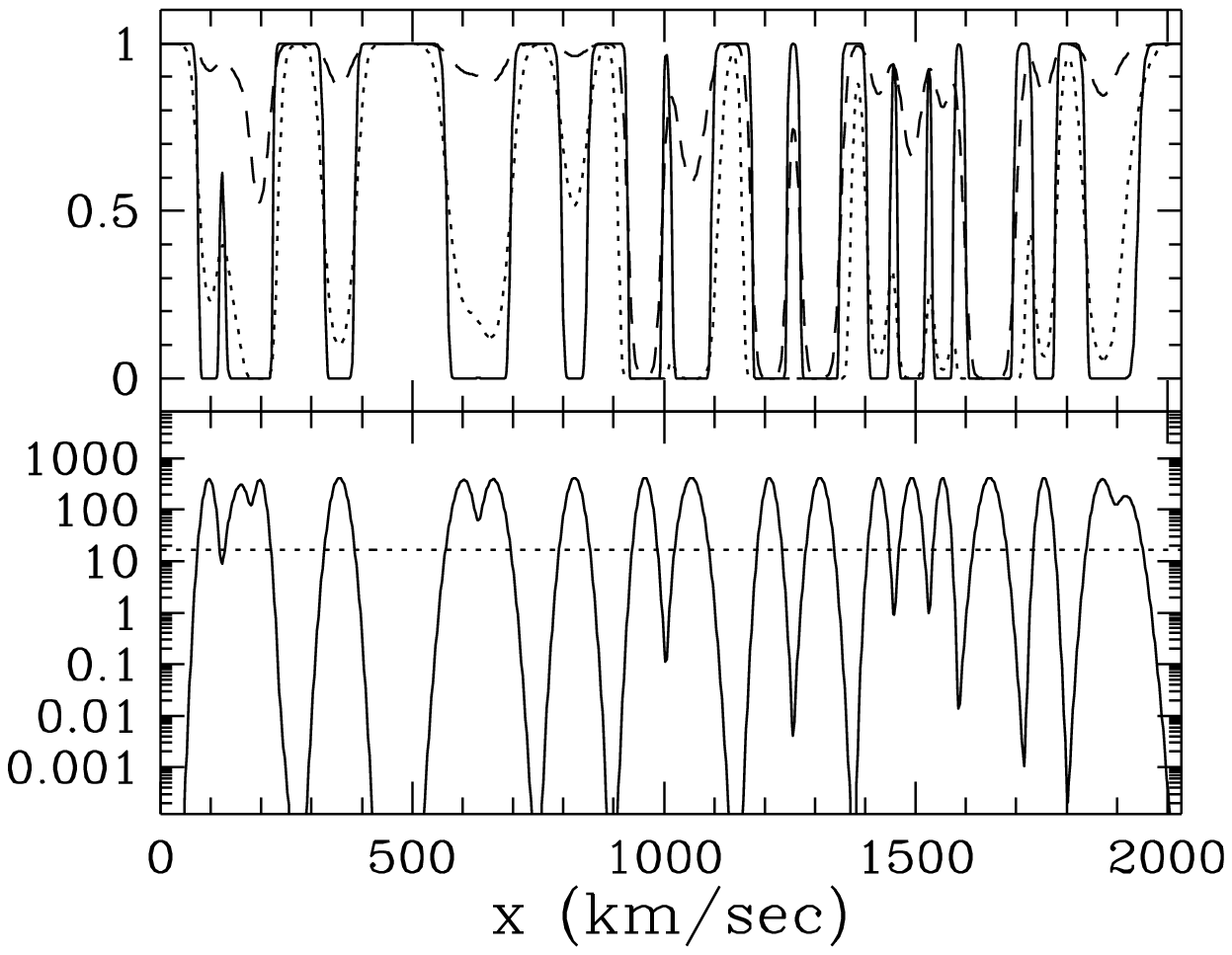}
\caption[junk]{\label{h1vsz}
Two spectra generated from a model of randomly distributed,
Voigt-profile lines.  Dashed lines in the top panels show
transmitted flux in the HI \lya\ spectrum.
Solid and dotted lines show HeII transmission assuming
thermal broadening and turbulent broadening, respectively.
The spectra are normalized to produce the observed values of
$\btauh$ and $\btauhe$.
The bottom panels show the ratio $\tauhe/\tauh$.
This is constant for the turbulent broadening model
(dotted line) because each line's HI and HeII $b$-parameters are 
identical.  In the thermal broadening model (solid line), 
the ratio varies strongly because of the factor of 2 difference
in $b$-parameters.
}
\end{figure}

Figure~7 shows two examples of the line model spectra.
Comparing to Figure~4, we see that while the HI spectra look qualitatively
similar to those produced by the cosmological simulations, the
HeII spectra look quite different --- they are sharply corrugated,
with many saturated regions in each spectrum.  
The simulations reproduce the observed values of
$\btauh$ and $\btauhe$ at $z=2.33$
simultaneously if $\Gamma_{\rm HI}/\Gamma_{\rm HeII} \approx 100$, as
implied by the HM background spectrum.
Our line models reproduce $\btauh$ and $\btauhe$ by construction, but
the required UV background spectrum is much softer,
with $\Gamma_{\rm HI}/\Gamma_{\rm HeII} \sim 200$ for the turbulently
broadened model and $\sim2500$ for the thermally broadened model.
If the minimum column density is pushed far below $10^{12}\;\cm^{-2}$,
then the qualitative properties of the line model become closer to
those of the simulations, since the weak lines overlap to produce a
fluctuating background that gives rise to much of the HeII absorption.
Whether the underlying physical picture approaches that of the 
simulations depends on how one envisions the absorbers themselves.
In a discrete cloud model, the absorption arises in physically distinct
objects whose wings overlap in frequency space because of line broadening,
but in the cosmological simulations the absorption arises in a
smoothly fluctuating, continuous IGM.

\subsection{Statistical analysis of transmission and optical depth}

We now turn to statistical measures that quantify properties 
of the absorbing gas in the various models that we have introduced.
These statistical predictions can be used to test and differentiate
these models.

\begin{figure}
\vspace{6.9cm}
\includegraphics{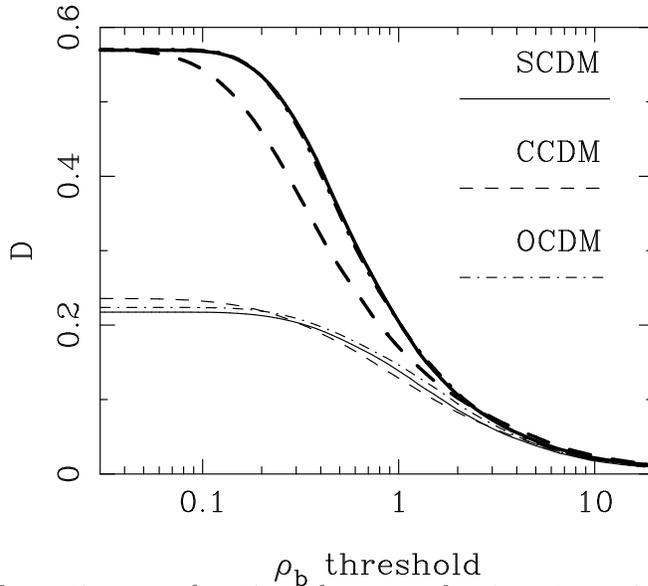}
\caption[junk]{\label{rhothresh}
Mean absorption as a function of gas overdensity, at $z=2.33$.
$D = \langle 1-F \rangle$ is the mean flux decrement in the spectrum
after the contribution of gas with density
below a threshold value of $\rho_{b}$ (in units of the mean baryon
density) is eliminated.  Thick lines show results for HeII,
thin lines for HI.  The SCDM, CCDM, and OCDM models are represented
by solid, dashed, and dot-dashed lines, respectively.
}
\end{figure}

Figure~8 demonstrates a point suggested qualitatively earlier:
most of the HeII absorption in the cosmological simulations arises
in low density regions.  We compute the mean flux decrement,
$D = \langle 1-F \rangle$, after setting to zero the
absorption caused by gas with baryon density below a specified threshold.
The flux decrement is plotted as function of 
the $\rho_{b}$ threshold (in units of the mean baryon density).
Regions with overdensity $\rho_b/\bar{\rho_b} < 2$ account for
roughly 80\% of the HeII flux decrement.
This density regime should be amenable to approximate analytic treatments ---
though not necessarily to linear perturbation theory {\it per se} ---
so analytic methods like that of 
Reisenegger \& Miralda-Escud\'{e} (1995; see numerical tests in
Miralda-Escud\'{e} $\etal$ 1996) should provide useful
guides to the predictions of HeII absorption in CDM-like models.
We see from Figure~8 that the SCDM and OCDM models, which have similar
mass fluctuation amplitudes, produce their HeII
absorption in almost identical density regimes.
In the higher amplitude, CCDM model, noticeably less
absorption arises in moderately underdense regions. 
As discussed in \S 3.1, more material in the CCDM model has flowed
out of ``voids'' into relatively dense objects, which produce saturated
HeII absorption.  If the intensity of the UV background were held constant,
this model would produce substantially less HeII absorption than the other 
two, as shown by the thin lines in Figure~3.  We have rescaled the 
background intensity so that all three models produce the same mean
absorption, but the greater emptiness of voids in the CCDM model
remains evident in the density distribution of the absorbing material.

When the same analysis is applied to HI absorption (thin lines in 
Figure~8), we see a shift towards higher density regions.
The density regime $\rho_b/\bar{\rho_b} < 2$ accounts for
$\sim 55\%$ of the HI flux decrement, compared to 80\% for HeII.
Regions below the mean density produce $\sim 65$\% of the HeII absorption
but only $\sim 35$\% of the HI absorption. 
Small differences between the models appear mainly because
scaling the UV background to match $\btauhe$ does not
give exactly the same $\btauh$ in each case.

\begin{figure}
\vspace{10.9cm}
\includegraphics{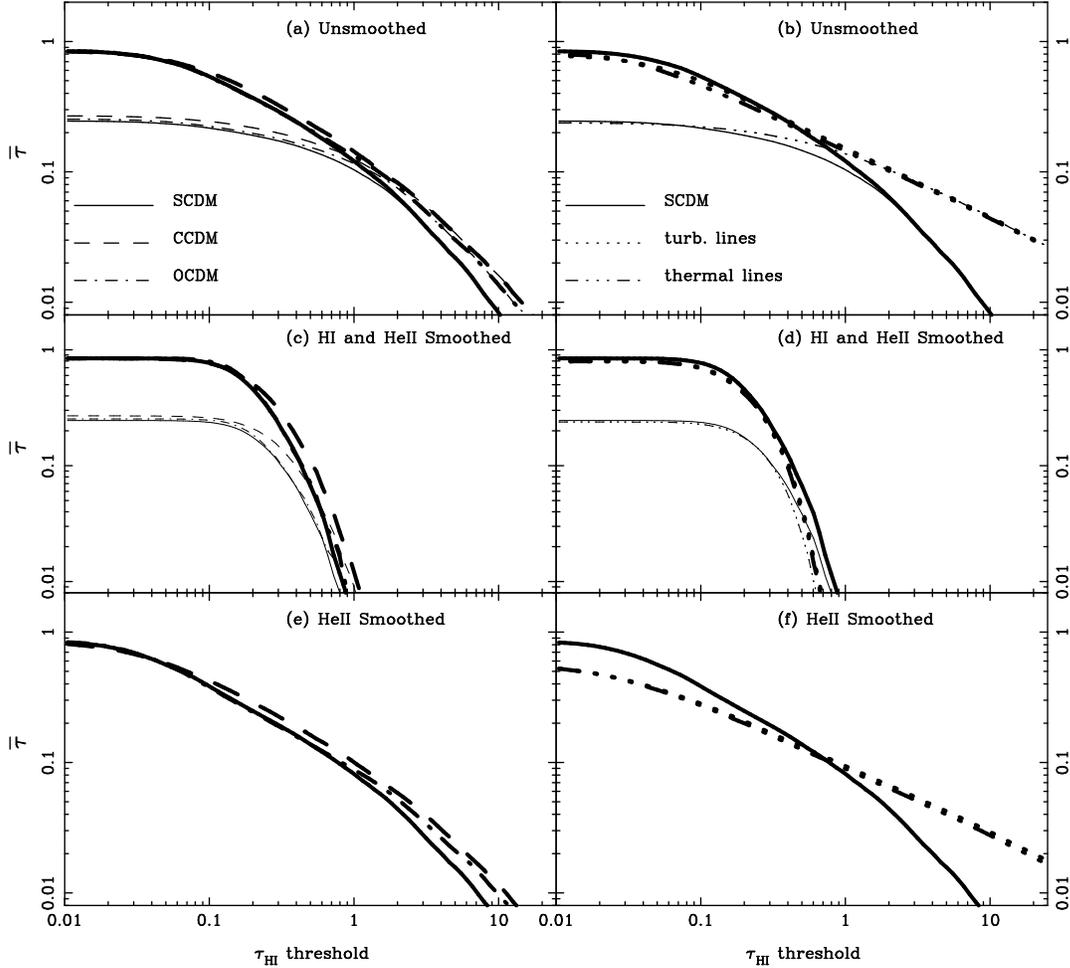}
\caption[junk]{\label{thresh}
Effective mean optical depth as a function of $\tauh$ threshold, at $z=2.33$,
analogous to Fig.\ 8.  The value of 
$\overline{\tau}=-\log_{e}\langle F \rangle$
is calculated after setting $F=1$ for points in the spectrum that
have optical depth $\tau_{\rm HI}$ below the threshold value on the
horizontal axis.  Thick lines show $\btauhe$ and thin lines $\btauh$.
Left hand panels compare the three CDM models; right hand panels
compare the SCDM model to the line models with turbulent and thermal 
broadening.  HI curves are identical for the two line models, since
they assume identical HI line populations.
Top panels show results for unsmoothed spectra.
Middle panels show results after HeII and HI spectra are smoothed
with a Gaussian filter of FWHM $\Delta v= 890\;\kms $,
corresponding to 3$\AA$ in the HeII spectrum. 
Bottom panels show results when the HeII spectra are smoothed but 
the HI spectra are not.  Only HeII curves appear in the bottom
panels, since the HI curves would be identical to those in the
top panels.
}
\end{figure}

The baryon density $\rho_b$ is not directly observable, but
it is well correlated with the HI optical depth, which is observable.
The upper left panel of Figure~9 is analogous to Figure~8,
except that we use a threshold in $\tauh$ instead of a threshold in $\rho_b$,
and we convert the mean flux decrement to an equivalent $\overline{\tau}$
in order to improve the dynamic range of the plot.
We set the absorption of regions with $\tauh$ below 
the threshold to zero before computing the mean decrement, $D$,
and $\overline{\tau} = -{\rm log}_e (1-D)$.
At high values of $\tauh$, the HI curves (thin lines) and HeII curves
(thick lines) for each model converge, since saturated regions completely
absorb both HI and HeII.
However, much of the HeII absorption is associated with regions that have
low HI optical depth.  For example, half of the contribution to $\btauhe$
comes from regions with $\tau_{\rm HI} < 0.15$. 
Only $\sim 15$\% of $\btauh$ is produced by these regions.
Figure~2 demonstrates that low density regions are revealed more clearly
by HeII absorption than by HI absorption.  Figure~9 demonstrates
that these regions are in fact responsible for much of the HeII
opacity of the high redshift universe.

The upper right panel of Figure~9 compares the SCDM model 
to the two line models.
The thermal and turbulent models 
are based on identical HI line populations,
so their HI curves are identical, and their HeII curves are very similar.
Even in the line models, much of the HeII absorption occurs in
regions of low HI optical depth ---  weak lines and the wings of
strong lines.  Relative to the CDM simulations, however, both line 
models produce a much larger fraction 
of $\btauhe$ in saturated regions.
This difference corresponds to the visual impression one obtains by
comparing sample spectra (Figures~4 and~7).

The DKZ data have a spectral resolution of approximately $3\AA$
(Gaussian FWHM), so the predictions in the top panels of Figure~9
cannot be compared directly to the DKZ observations.  The middle panels
show the same analysis after the model HI and HeII spectra
have been convolved with a Gaussian filter of $890\;\kms$ FWHM,
equivalent to $3\AA$ for HeII at $z=2.33$.  
Regrettably, this smoothing eliminates most of the difference
between the CDM models and the line models.
The sense of this difference is reversed relative to the full
resolution case, with the CDM models producing a slightly larger 
fraction of their absorption at high values of (smoothed) $\tauh$.
This change probably reflects the large scale clustering that
is present in the CDM models but not in the line models.
Among the CDM models themselves, it is the more strongly clustered,
CCDM model that produces the largest fraction of its absorption
at high $\tauh$.

The bottom panels of Figure~9 repeat the optical depth analysis
using smoothed HeII spectra but unsmoothed HI spectra, in
recognition of the fact that the best ground-based spectra
are able to resolve even the narrowest observed HI \lya\ features.
This version of the analysis preserves the clear distinction
between the CDM models and the line models visible
in the upper panels.  Indeed, smoothing the line models amplifies
the difference between them and the CDM models, since a significant
fraction of the line models' smoothed HeII flux decrement arises in 
``inter-line'' regions where the HI optical depth is extremely low.
While a realistic comparison to observations will have to contend
with noise in the HeII and (to a lesser extent) HI data, the 
distinction between the line models and the CDM models is likely
to be observable.  The distinctions among the CDM models themselves
are more subtle.
 
\begin{figure}
\vspace{10.9cm}
\includegraphics{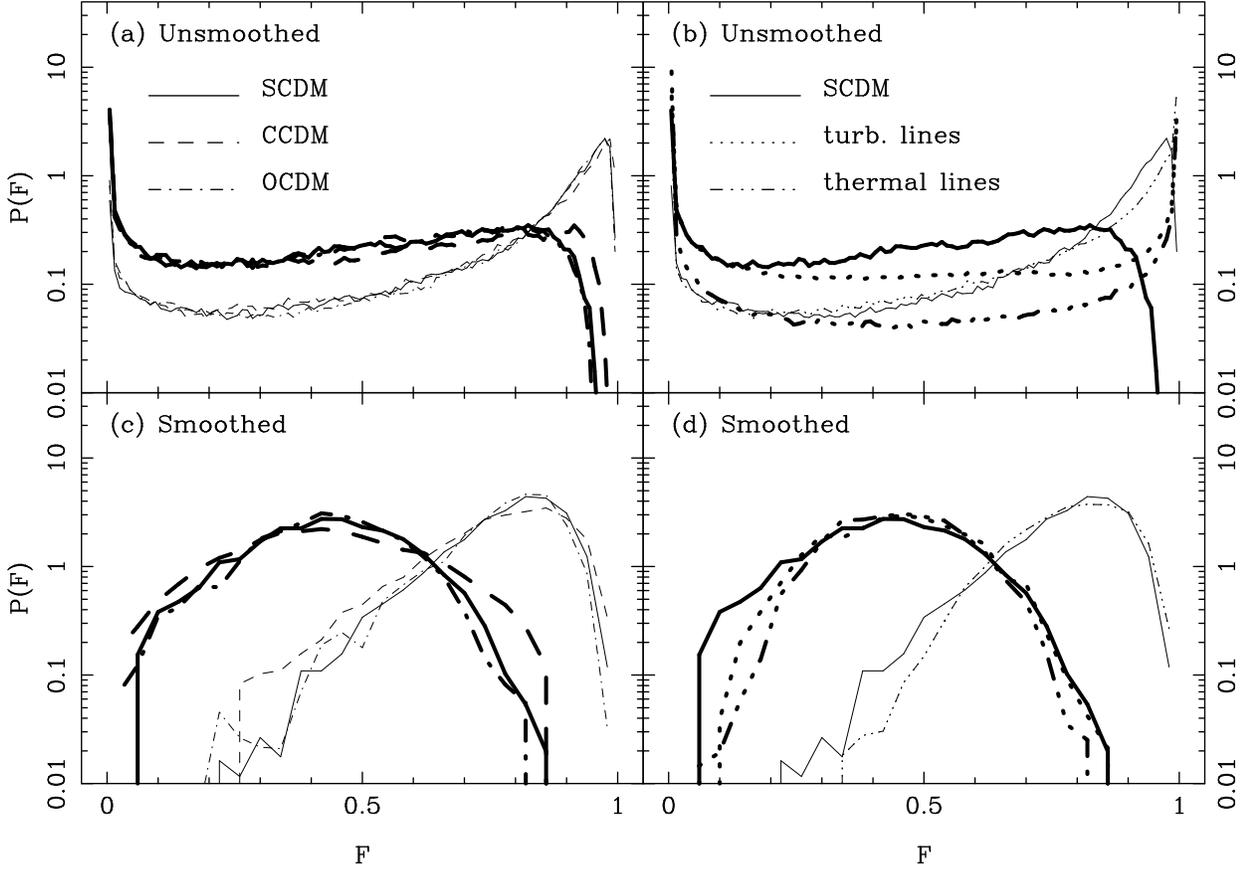}
\caption[junk]{\label{pdf}
The distribution function of transmitted flux for z=2.33.
$P(F)dF$ is the probability that a randomly selected point in the
spectrum has transmitted flux in the range $F \longrightarrow F+dF$.
As in Fig.~9, thick lines show HeII and thin lines HI,
left hand panels compare the CDM models, and right hand panels compare
SCDM to the two line models.
Top panels show results for unsmoothed spectra.
Bottom panels show results after the spectra are smoothed with a
Gaussian of FWHM $\Delta v=890\;\kms$.
}
\end{figure}

\begin{figure}
\vspace{10.9cm}
\includegraphics{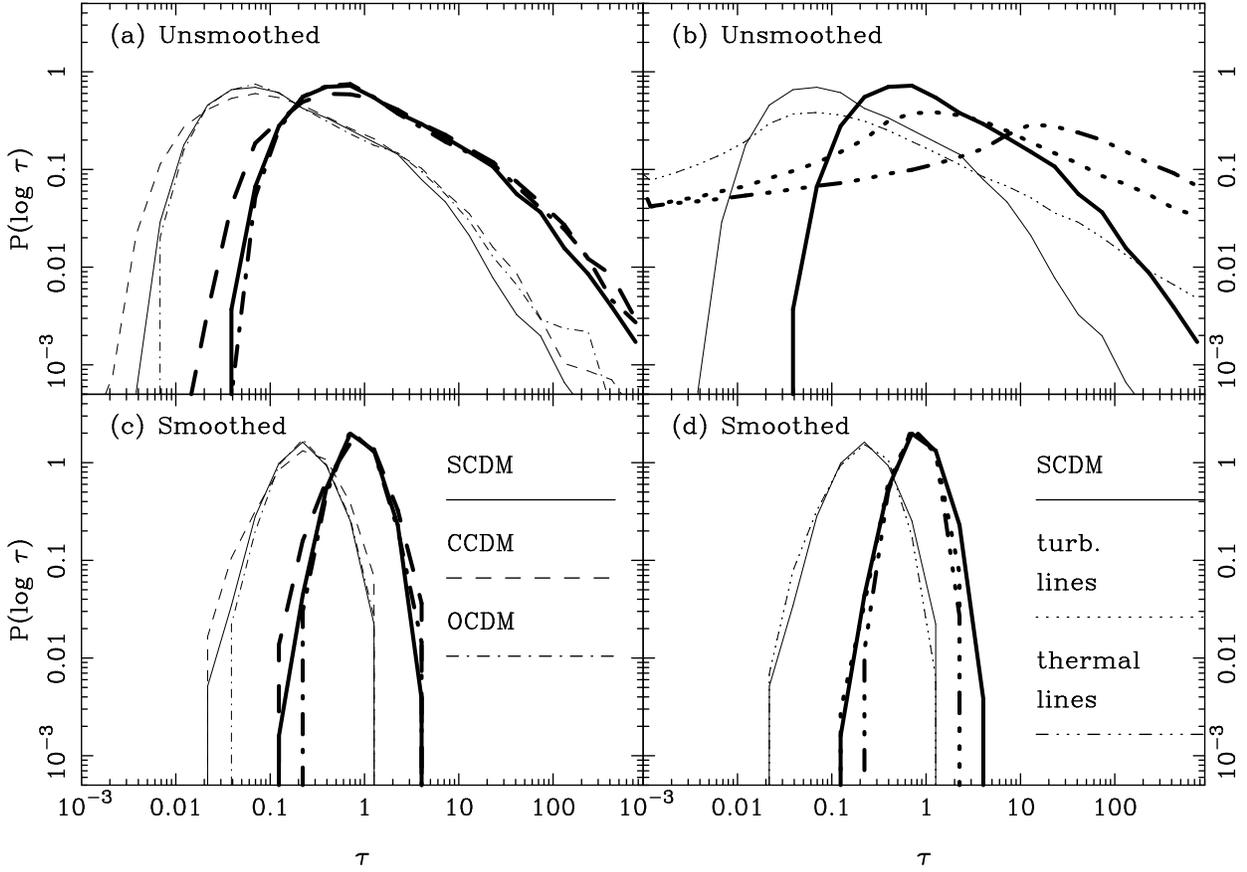}
\caption[junk]{\label{pdf}
The distribution function of optical depth in logarithmic bins for z=2.33.
$P(\log\tau)d\log\tau$ is the probability that a randomly selected point
on the spectrum has the log of the optical depth in the range 
$\log\tau \longrightarrow \log\tau + d\log\tau$.
The significance of panels and line types is the same as in Fig.~10.
}
\end{figure}

Figure~10 shows the distribution function of transmitted flux $F$;
$P(F)dF$ is the probability that a randomly selected point in the
spectrum has transmitted flux in the infinitesimal range 
$F \longrightarrow F+dF$.  Thick and thin lines represent HeII and
HI results, respectively.  Left hand panels compare the three CDM models,
and right hand panels compare the SCDM model to the line models.
Upper panels are based on unsmoothed spectra, lower panels
on spectra smoothed over $890\;\kms$.
Figure~11 shows the distribution function of the log of the optical
depth normalized so that $\int P(\log\tau)) d\log(\tau)=1$.
While the information is equivalent in principle to that in Figure~10,
the optical depth plot reveals tails of the distribution function more clearly.

The distribution functions of the three CDM models are rather similar,
though they are somewhat broader for the CCDM model because of its
higher clustering amplitude.  At full resolution, the HeII distribution
functions of the CDM models are radically different from those of 
the line models, especially at low optical depths ($F \approx 1$).
The line models have gaps in their spectra where the HeII absorption
is extremely low, but the CDM models do not.
The resulting difference in $P(F)$ --- an upturn at $F \approx 1$ for
the line models but a downturn for the CDM models --- also appears
in the HI spectra, but it is less dramatic, and the downturn occurs
at such low optical depth that it might be masked by errors in continuum
fitting.  The line models also have smaller fractions of the spectrum
with transmission in the middle range $0.05 < F < 0.9$.
The $P(\log\tau)$ distributions of the line models are much broader
than those of the CDM models, reflecting the same trends seen in $P(F)$.
Unfortunately, the strong differences in the HeII spectra are 
mostly lost when the spectra are smoothed over $3\AA$.  
The higher resolution spectra obtainable with HST may be better 
suited to detecting or ruling out the gaps in HeII absorption
predicted by the line models, even though HST can only probe
redshifts $z > 3$.  Our simulations tend to underestimate the
widths of $P(F)$ and $P(\log\tau)$ for the smoothed spectra because the 
$890\;\kms$ filter width is a substantial fraction of our box size
($2028\;\kms$ at $z=2.33$), and because we do not include spatial
fluctuations in the intensity of the ionizing background
(Zuo 1992; Fardal \& Shull 1993), which could be a significant 
source of additional fluctuations in the HeII opacity.
A comparison of CDM simulations to ground-based,
HI data using the cumulative form of $P(F)$ is presented in
Rauch et al.\ (1997) and Croft et al. (1997).  For further discussion of this
statistic, see Miralda-Escud\'e et al.\ (1996).

\begin{figure}
\vspace{10.9cm}
\includegraphics{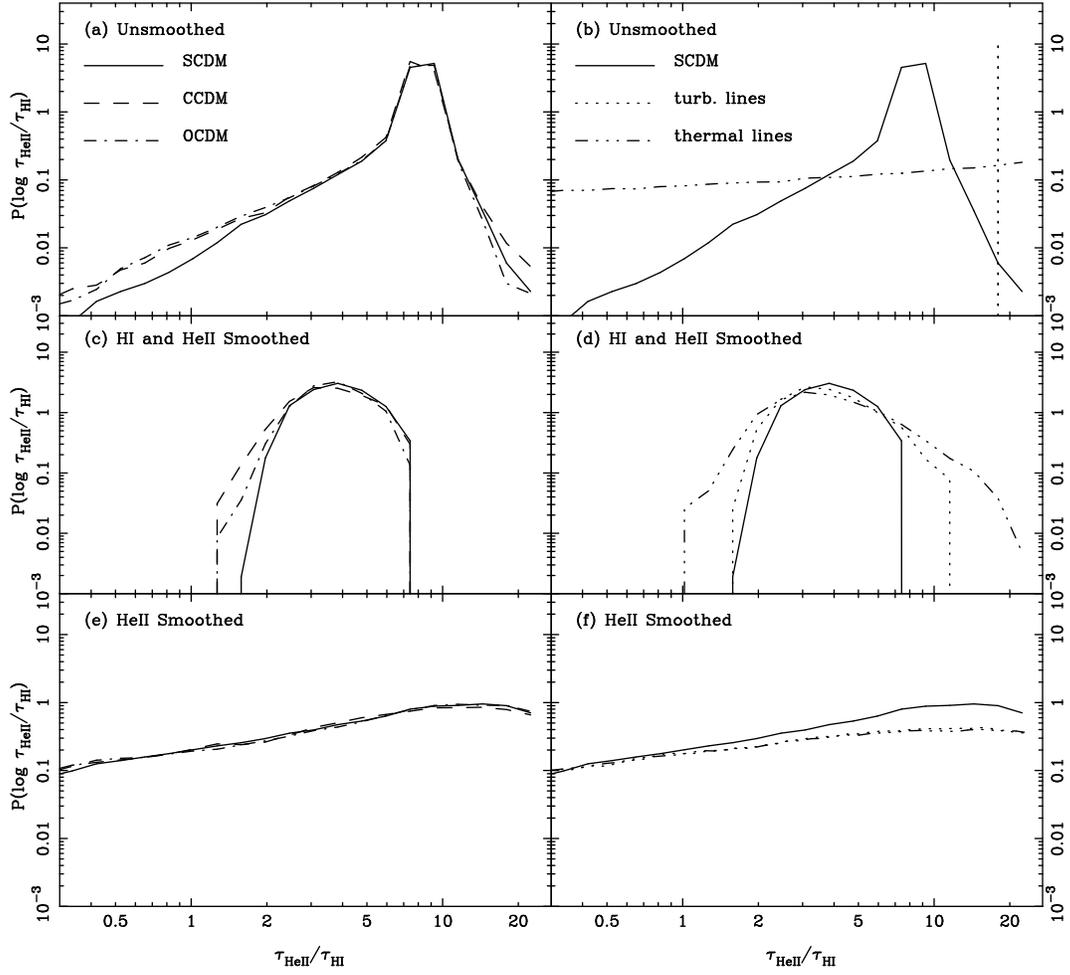}
\caption[junk]{\label{ratio}
Distribution function of the log of the optical depth ratio, 
$\log(\tauhe/\tauh)$, at $z=2.33$.
Layout is similar to that of Figure~9.
}
\end{figure}

Figure~12 shows the distribution function of the log of the optical depth
ratio,
$\log(\tauhe/\tauh)$.  As in Figure~9, the top panels show results for
unsmoothed spectra, the middle panels show results with both spectra
smoothed over $890\;\kms$, and the bottom panels show results using
smoothed spectra for $\tauhe$ but unsmoothed spectra for $\tauh$.
Starting with the top left panel, we see that the three CDM models
predict similar distributions of $\log(\tauhe/\tauh)$, all
peaked around $\tauhe/\tauh \approx 8$.
The extended tails towards low $\tauhe/\tauh$ are caused by thermal
broadening, and the shorter tails towards high $\tauhe/\tauh$
are caused by collisional ionization, as discussed in \S 3.3.
In the smoothed spectra (middle left panel), the CCDM and OCDM models
have broader distributions than the SCDM model, though the regime
where this difference is strong is well below the peak of the distribution,
and it may therefore be difficult to probe observationally.
For smoothed HeII but unsmoothed HI (bottom left panel), the distributions
become extremely broad and virtually identical.  
In particular, there are strong tails towards
low $\tauhe/\tauh$ caused by strong HI features whose HeII counterparts
have been reduced by smoothing.  The distributions also extend to high
$\tauhe/\tauh$ because the smoothed HeII optical depth is rarely less
than 0.2, while the optical depth in the unsmoothed HI spectra can
be 0.01 or smaller (see Figure~11).  

Right hand panels of Figure~12 compare the SCDM model to the line models.
Without smoothing, the turbulent and thermal broadening models define
two extremes, both very different from the CDM results.
By construction, the turbulent broadening model has a constant ratio
of $\tauhe$ to $\tauh$, so its distribution is a $\delta$-function
at $\tauhe/\tauh=17$.  
In the thermal broadening model, however, the HI and HeII $b$-parameters
differ by a factor of two, so in every line the ratio $\tauhe/\tauh$ is
high in the core and low in the wings.
When the HeII and HI spectra are smoothed, the line model distributions
are broader than the CDM distributions, especially in the direction of
high $\tauhe/\tauh$.  However, the strong differences again occur well
below the peak of the distribution.  Dividing smoothed HeII optical depths
by unsmoothed HI optical depths again yields a broad distribution function,
but in this case the CDM results lie significantly above the line model
results at the peak of the distribution, providing a clear distinction
between the two sets of models.  As in Figure~9,
the difference reflects the presence of gaps in the line model HI spectra.

Figures~9 and~12 show two useful 1-dimensional summaries
of information contained in the 2-dimensional, joint distribution of
HI and HeII optical depths (Figure~5).  Figure~11 shows the 
projections of this distribution along the HI and HeII axes.
There are other potential cuts through this joint distribution,
or one could use the full 2-dimensional distribution itself to
distinguish between models.

Our cosmological simulations have finite resolution, as discussed in \S 2.
The gravitational softening length, 10 kpc comoving, is much smaller than
the typical scale of \lya\ absorbers, but we do not adequately resolve baryonic
structures less massive than $\sim 32$ SPH particles 
($4.65 \times 10^9 M_\odot$ for SCDM and CCDM, $2.11 \times 10^9 M_\odot$
for OCDM), and we cannot represent the initial fluctuation spectrum
for wavelengths smaller than twice the initial particle grid spacing,
$\lambda=350\; h^{-1}\;$kpc comoving.
If we simulated the same volumes with much larger particle numbers,
we would therefore expect some differences of detail for the statistical
results illustrated in this section.  However, the regions of the IGM
that dominate HeII and HI absorption are low density and fairly smooth,
and they lie outside the density/temperature regime in which cooling
instabilities play an important role, so we do not expect major
changes in our results to appear at higher resolution.  
We cannot firmly estimate
the quantitative impact of resolution effects until we are able to 
perform simulations with more particles, which will be possible with 
a parallel TreeSPH code now under development
(Dav\'{e}, Dubinski \& Hernquist 1997).  We expect that the main
qualitative effect of increased resolution on the simulated IGM will be a 
larger amount of low-amplitude substructure in underdense regions.
Simulations with a factor of eight {\it fewer} particles than the
ones used here yield
similar physical properties for the IGM and a similar column density
distribution for HI \lya\ lines with $N_{\rm HI} \la 10^{15} \cm^{-2}$,
but many fewer systems at higher column densities,
where radiative cooling becomes important (Miralda-Escud\'e et al., in
preparation).

The principal physical uncertainty in our simulations is the impact of
reionization on gas temperatures.  As pointed out by 
Miralda-Escud\'e \& Rees (1994), the energy injection of one photoelectron
per proton during reionization can heat the IGM
to several$\;\times\;10^4$K, {\it if} reionization occurs rapidly
enough that this energy is not dissipated by collisional line cooling.
Our equilibrium treatment of photoionization suppresses this heating
because we set neutral fractions to low values as soon as the
ionizing background switches on, without altering gas temperatures.
This treatment is equivalent to assuming that reionization occurred
slowly (with consequent radiative cooling) or at high redshift
(with consequent heat losses to Compton and adiabatic cooling).
If we instead assumed that the IGM at $z \sim 2-3.5$ retained significant
heat from reionization, then the temperatures of the unshocked or
weakly shocked gas would be higher and less dependent on
density.  Recall that for the gas that produces most of the absorption,
the HI and HeII optical depths are proportional to 
$\rho_b T^{-0.7} \Gamma^{-1}$.  With hotter gas, the required values
of $\fh$ and $\fhe$ would therefore be higher, i.e., our models
would require a less intense UV background or a higher baryon density
in order to match the observed $\btauh$ and $\btauhe$.
For a fixed reionization history, the $(\Omega_b h^2)^{1.7}$ 
dependence in equation~(8) would be closer to $(\Omega_b h^2)^2$, since
the effect of reionization heating on gas temperatures does not depend 
on $\Omega_b$ (though the subsequent photoionization heating does).
The magnitude of these effects depends on the reionization model;
plausible models could increase $\fh$ and $\fhe$ by $25-50\%$
(see the discussion by Hui \& Gnedin 1997).
However, once the simulations were normalized to the observed $\btauh$
and $\btauhe$, we would expect the statistical properties of the 
absorption to be very similar to those computed here, with small
changes reflecting the weaker dependence of optical depth on density
and the larger degree of thermal broadening. Plausible changes
in the reionization history all go in the direction of increasing rather 
than decreasing IGM temperatures, so they only tend to raise the
lower bounds to $\Omega_{b}$ discussed in \S 3.2. IGM temperatures and 
corresponding $\Omega_{b}$ limits could be lower if the UV background 
spectrum is substantially softer than the HM spectrum, so that the mean
residual photoelectron energy is lower.

\section{Conclusions}
Many previous discussions of HeII absorption
have focused on distinguishing ``line blanketing'' in the \lya\
forest from ``Gunn-Peterson'' absorption by the IGM.
Our cosmological simulations undermine the premise of this effort,
for they suggest that HeII absorption and the low column density \lya\ 
forest {\it both} arise in diffuse, smoothly fluctuating, intergalactic gas.
Local maxima in the optical depth can be identified as lines,
but individual features do not, as a rule, correspond to compact structures
that are sharply separated from their environment.
While the \lya\ forest and the HeII flux decrement are both 
manifestations of the IGM, the high ratio of HeII ions to HI atoms
does lead to an important systematic difference 
between helium and hydrogen absorption:
HeII absorption is stronger in the mean,
and much of it arises in underdense regions that have low HI optical depth.
The general picture of the IGM presented here is similar to that
in the other numerical simulation papers cited in the introduction, and it
has much in common with the semi-analytic models developed by Bi (1993),
Bi, Ge, \& Fang (1995), Bi \& Davidsen (1997), and 
Hui, Gnedin, \& Zhang (1997).  These semi-analytic models
lead to similar qualitative conclusions 
about the properties of the gas producing HeII absorption
(Davidsen, private communication; Davidsen et al., in preparation).

Some of the more specific conclusions from our analysis are as follows:
\\
(1) The CDM models account for the observed relative values of $\btauh$
(from PRS) and $\btauhe$ (from DKZ and HAR) if the UV background has
the spectral shape predicted by HM.
Large changes in the spectral shape (or in the observational estimates)
would spoil this agreement.
\\
(2) These models account for the observed absolute values of
$\btauh$ and $\btauhe$ only if (a) the overall intensity of the background
is lower than predicted by HM, or (b) the baryon density is higher than
our assumed value of $\Omega_b h^2=0.0125$.
If we set the background intensity equal to the HM value,
then the SCDM and OCDM models require $\Omega_b h^2 \approx 0.023$,
and the CCDM model requires $\Omega_b h^2 \approx 0.038$.
We have analyzed one SCDM simulation with a higher baryon density,
$\Omega_b h^2=0.03125$.
If $\Gamma_{\rm HeII}$ is held fixed, then this model produces stronger
HeII absorption than the original SCDM model, as expected.
Once both models are
normalized to produce the same $\btauhe$, the statistical properties
of their HeII absorption (e.g., the measures considered in \S 3.4)
are virtually identical.
\\
(3) The CDM models naturally explain the observed evolution of $\btauhe$,
in particular the factor of two drop in $\btauhe$ between 
$z\approx 3.3$ (HAR) and $z\approx 2.4$ (DKZ),
provided $\Gamma_{\rm HeII}$ evolves at the rate calculated by HM.
Change in the UV background does not play a major role in this evolution ---
$\Gamma_{\rm HeII}$ grows by only 20\% between $z=3.3$ and $z=2.4$.
The strong evolution of $\btauhe$ is instead driven by the expansion of 
the universe, which lowers gas densities and spreads absorbing material over
larger frequency ranges.
The high HeII optical depth measured by HAR does not imply that
HeII reionization occurred at $z<3.3$.
If the HAR upper limit of $\btauhe<3.0$ outside the Q0302-003 ionization
zone is correct, then the photoionizing background
along this line of sight cannot be much softer than HM predict.
\\
(4) Most of the HeII opacity is produced by diffuse gas that follows a
well defined relation between temperature and density. 
This relation has its origin in
the competition between photoionization heating and adiabatic cooling.
For gas that lies on this temperature-density relation, the HeII optical
depth is a simple function of density, $\tau_{\rm HeII}\propto \rho_{b}^{1.6}$.
To a first approximation, one can regard a HeII (or HI) absorption spectrum as
a non-linear map of the gas density along the line of sight.
\\
(5) A significant fraction of the HeII absorption arises in regions
with density contrast $\delta\leq -0.5$, for which linear perturbation
theory will give inaccurate results.  Analytic methods that treat
this density regime more accurately, such as
the Modified Zeldovich Approximation of Reisenegger \& Miralda-Escud\'{e}
(1995), may provide useful guides to the behavior of HeII absorption
in cosmological models, especially in light of point (4) above.
\\
(6) The three CDM models that we have investigated predict similar
statistical properties of the HeII absorption (e.g., distribution
functions of $\tauhe$ and $\tauhe/\tauh$), once they are normalized
to produce the same mean absorption.  The CCDM model predicts somewhat
broader distribution functions than the SCDM or OCDM models
because of its higher mass fluctuation 
amplitude.  Because this model has emptier voids, it also
requires a lower UV background intensity and/or higher $\Omega_b h^2$ 
to reproduce the observed $\btauhe$, and it requires a slightly softer 
background spectrum to account simultaneously for $\btauh$ and $\btauhe$.
On the scales of our simulation, the three CDM models have power spectra
of similar shapes, and their rms fluctuation amplitudes differ by
less than a factor of two, so we do not yet know
how the absorption results might change for much steeper or shallower
power spectra or for models that have very different fluctuation
amplitudes on these scales.
\\
(7) The CDM models have very different HeII absorption properties from
our ``line models,'' which assume that the HeII absorption is produced
by randomly distributed, Voigt-profile lines with 
$dN/dN_{\rm HI} \propto N_{\rm HI}^{-1.5}$ and a lower cutoff at
$N_{\rm HI} = 10^{12}\;\cm^{-2}$.  In particular, the line models
have gaps in which the HeII absorption is very low,
and a larger fraction of their HeII absorption arises in regions
of high HI optical depth.  The fluctuating IGM of the CDM models
produces fluctuating HeII absorption, but absorption-free
regions ($\tauhe \la 0.05$) are very rare.
The line models require a softer UV background spectrum 
in order to produce the observed $\btauhe$, especially if the lines are 
thermally broadened.  The statistical differences between the line models and
the CDM models are greatly reduced when the spectra are smoothed, so
for distinguishing these scenarios it is desirable to use HeII
spectra with the highest resolution practical.
If the minimum column density
in the line models is pushed well below $10^{12}\;\cm^{-2}$, the 
differences from the CDM models are less striking, as weak lines
overlap to produce a continuous, fluctuating background.
\\
(8) About half of the contribution to $\btauhe$ comes from regions that
have $\tauh < 0.15$.  In the CDM models, most of the absorption in this
regime is produced by a smoothly fluctuating IGM rather than the wings
of strong absorption lines.  The absorption is not uniform, but because
the HI optical depth is low and the variations are gentle, much of the
corresponding HI absorption could be inadvertently
removed from optical quasar spectra in the process of continuum fitting.  

The CDM simulations provide a model of the high redshift IGM.
Once it is normalized to the observed values of $\btauhe$ and $\btauh$,
this model yields a number of testable predictions.
First, the shape of the UV background spectrum should be close to
that predicted by HM, with $\Gamma_{\rm HI}/\Gamma_{\rm HeII} \approx 100$,
for $2 \lesssim z \lesssim 3.3$.  It is difficult to measure the background
shape precisely independent of an IGM model, but tests using metal
line ratios (Songaila \& Cowie 1995) or comparisons of the HI and HeII
proximity effects (see Zheng \& Davidsen 1995) might be able to 
identify strong departures from the HM spectral shape.
If one adopts the intensity of the HM background as a lower
limit, then our models also predict that the baryon density exceeds
the ``conventional'' big bang nucleosynthesis estimate of $\Omega_b h^2=0.0125$
by a factor $\sim 1.5-3$ (see Rauch et al. 1997 and
Weinberg et al. 1997 for further
discussion).  The most direct and quantitative predictions of this IGM 
scenario are statistical properties of the absorption such as those
shown in \S 3.4.  In particular, observational analyses can test the
predictions that a large fraction of the HeII opacity arises in regions
of low HI optical depth and that there are few regions where $\tauhe$
itself is very low.  The DKZ spectrum offers the largest redshift
range for statistical analyses of HeII absorption.  However, the
distinctive features of this IGM model are most stringently tested at
high spectral resolution, so HST observations at $z>3$ can also play
an important role.  An anecdotal but significant argument in
favor of this IGM model is provided by HAR, who show that strong HeII
absorption ($\tauhe > 1.3$) persists in a region where a Keck HIRES
spectrum reveals no HI \lya\ lines down to a detection threshold $\tauh =0.05$.

Careful comparison between the simulations and HeII observations should
include realistic treatments of noise, instrumental resolution, the
redshift range of the observations, and so forth.  The three variants
of the CDM model considered in this paper yield fairly similar 
predictions, and given the inevitable limitations of HeII measurements
with existing instruments, we anticipate that HI observations will be
a more powerful tool for distinguishing among them.  However, HeII
absorption provides a vital test of this general scenario for high
redshift structure, checking one of its basic predictions
in a regime that is almost inaccessible
to other methods.  In theories of structure formation like those 
studied here, underdense regions of the universe harbor a substantial
amount of diffuse baryonic material.  If these theories are even roughly
correct, then the recent studies of HeII absorption in quasar spectra
have detected this material at high significance.  More detailed analyses
of existing absorption data can test this claim.  Future instruments
that can observe HeII absorption at high spectral resolution
could yield a precise view of structure in the sparsest
regions of the high-redshift universe.

\bigskip
 
{\large\bf Acknowledgements}
 
\bigskip

We acknowledge inspiring talks by Peter Jakobsen and Arthur Davidsen
on HeII absorption observations at an STSci workshop on QSO absorption
lines in June, 1995.  We thank Arthur Davidsen, Craig Hogan,
Jordi Miralda-Escud\'e, and Andreas Reisenegger
for stimulating and informative discussions about HeII absorption.
We thank Piero Madau for helpful discussions on the UV background
and for providing the HM background spectrum
in convenient numerical form.
We also thank Jordi Miralda-Escud\'e for providing us with the
computer program used to generate the line model spectra.
The simulations were performed at the San Diego Supercomputer Center.
This work was supported by NASA Astrophysical Theory Grants
NAG5-2864, NAG5-3111, NAGW-2422, NAG5-2793,
by NASA Long-Term Space Astrophysics Grant NAG5-3525,
by NASA HPCC/ESS grant NAG5-2213,
and by the NSF under grants ASC93-18185 and the Presidential
Faculty Fellows Program.

\bigskip 
 
\setlength{\parindent}{0mm}
{\bf REFERENCES} 
\bigskip
 
\def\refe {\par \hangindent=.7cm \hangafter=1 \noindent}
\def\apj { ApJ }
\def\mn { MNRAS }
\def\apl { ApJ}
\def\aa { A\&A}
\def\nat {Nature}

\refe Bajtlik, S., Duncan, R.C., Ostriker, J. P. 1988, \apj, 327, 570
 
\refe Bennett, C., L., Banday, A. J., Gorsk\'{i}, K. M., Hinshaw, G.,
Jackson, P., Keegstra, P., Kogut, A., Smoot, G. F., Wilkinson, D. T.,
Wright, E. L. 1996, \apl, 646, L1  

\refe Bi, H. 1993, \apj, 405, 479

\refe Bi, H., \& Davidsen, A. F. 1997, \apj, 479, 523

\refe Bi, H., Ge, J., \& Fang, L.-Z. 1995, \apj, 452, 90

\refe Carswell, R. F., Rauch, M., Weymann, R. J., Cooke, A.J., Webb, J.K.,
1994, \mn, 268, L1

\refe Cen, R., Miralda-Escud\'{e}, J., Ostriker, J.P., Rauch, M.,
1994, \apl, 437, L9 
 
\refe Croft, R. A. C., Weinberg, D. H., Hernquist, L. \& Katz, N.,  1997,
in  Proceedings of the 18th Texas Symposium on 
Relativistic Astrophysics, eds. Olinto, A., Frieman, J. \& Schramm, D.,
World Scientific), astro-ph 9701166

\refe Dav\'{e}, R., Dubinski, J., Hernquist, L., 1997,
New Astronomy,
submitted, astro-ph 9701113

\refe Dav\'{e}, R., Hernquist, L., Weinberg, D. H., Katz, N. 1997, \apj,
477, 21

\refe Davidsen, A.F., Kriss, G. A. \& Zheng, W., \nat, 380, 47 (DKZ)

\refe Efstathiou, G., Bond, J. R., \& White, S. D. M.,
1992, \mn, 258, 1p

\refe Fardal, M. A., \& Shull, M. J. 1993, \apj, 415, 524

\refe Giallongo, E., Cristiani, S., Dodorico, S., Fontana, A., Savaglio, S.,
1996, \apj, 466, 46
 
\refe Giroux, M.L., Fardal, M.A., \& Shull, J.M. 1995, \apj, 451, 477

\refe Gorski, K. M., Ratra, B., Sugiyama, N., \& Banday, A. J. 1995,
\apj, 444, L65

\refe Gunn, J. E. \& Peterson, B. A. 1965, \apj, 142, 1633
 
\refe Haardt, F. \& Madau, P. 1996, \apj, 461, 20 (HM)
 
\refe Hellsten, U., Dav\'{e}, R., Hernquist, L., \& Katz, N., 1997, \apj,
in press, astro-ph 9701043

\refe Hernquist, L. \& Katz, N. 1989, ApJS, 70, 419 
 
\refe Hernquist, L., Katz, N., Weinberg, D.H., Miralda-Escud\'{e}, J.,
1996, \apj, 457, L51 (HKWM)

\refe Hogan, C.J., 1997, in Proceedings of the 18th Texas Symposium on 
Relativistic Astrophysics, eds. Olinto, A., Frieman, J. \& Schramm, D.,
World Scientific).
 
\refe Hogan, C.J., Anderson, S.F. \& Rugers, M.H. 1996, \aj, in press (HAR)
 
\refe Hu, E.M., Kim, T.-S., Cowie, L.L., Songaila, A. \& Rauch, M. 1995,
AJ, 110, 1526

\refe Hui, L., \& Gnedin, 1996, \mn, submitted,  astro-ph/9612232

\refe Hui, L., Gnedin, N. Y., \& Zhang, Y. 1997, \apj, in press, 
astro-ph/9608157
 
\refe Jakobsen, P., Boksenberg, A., Deharveng, J. M.,
Greenfield, P., Jedrzejewski, R., Paresce, F. 1994, \nat,
370, 35 (JBDGJP)
 
\refe Katz, N., Weinberg, D.H. \& Hernquist, L. 1996, ApJS, 105, 19 (KWH)
 
\refe Lynds, C.R. 1971, \apl, 164, L73
 
\refe Madau, P., \& Meiksin, A. 1994, \apl, 433, L53
 
\refe Miralda-Escud\'{e}, J. 1993, \mn, 262, 273
 
\refe Miralda-Escud\'{e},J., Cen, R., Ostriker, J.P. \& Rauch, M.,
1996, \apj, 471, 582
 
\refe Miralda-Escud\'{e}, J., \& Ostriker, J. P. 1992, \apj, 392, 15

\refe Miralda-Escud\'{e}, J., \& Rees,  M. J. 1993, \mn, 260, 617

\refe Miralda-Escud\'{e}, J., \& Rees,  M. J. 1994, \mn, 266, 343
 
\refe Press, W., H., Rybicki, G. B., \& Schneider, D. P. 1993, \apj, 414, 64
(PRS)

\refe Rauch, M.,  Miralda-Escud\'{e}, J., Sargent, W. L. W., Barlow, T. H.,
Weinberg, D. H., Hernquist, L., Katz, N., Cen, R., Ostriker, J. P. 1997,
\apj, submitted, astro-ph 9612245
 
\refe Reisenegger, A. \& Miralda-Escud\'{e}, J. 1995, \apj, 449, 476

\refe Rugers, M., \& Hogan, C.J. 1996, \apl, 459, L1

\refe Rugers, M., \& Hogan, C.J. 1996b, \aj, 111, 2135
 
\refe Sargent, W.L.W., Young, P.J.,  Boksenberg, A., Tytler, D. 1980,
ApJS, 42, 41
 
\refe Songaila, A. \& Cowie, L.L. 1996, AJ, 112, 335
 
\refe Songaila, A. \& Cowie, L.L., Hogan, C.J.,
Rugers, M. 1994, \nat, 368, 599
 
\refe Songaila, A., Hu, E. M., \& Cowie, L.L. 1995, \nat, 375, 124

\refe Songaila, A., Wampler, E. J., \& Cowie, L.L. 1997, \nat, 385, 137

\refe Tytler, D., Fan, X. \& Burles, S. 1996, \nat, 281, 207

\refe Tytler, D., Burles, S., \& Kirkman, D., 1996,  \apj, submitted, astro-ph
9612121 

\refe Walker T. P., Steigman, G., Schramm, D. N., Olive, K. A., Kang, H. S.,
1991, \apj, 376, 51

\refe White, S. D. M., Efstathiou, G. P. \& Frenk, C. S. 1993, \mn, 262, 1023
 
\refe Weinberg, D.H., Hernquist, L., \& Katz, N. 1997, \apj, 477, 8

\refe Weinberg, D.H., Miralda-Escud\'{e}, J., Hernquist, L., \& Katz, N., 1997,
\apj, submitted, astro-ph 9701012  

\refe Zhang, Y., Anninos, P., Norman, M.L. 1995, \apl, 453, L57

\refe Zhang, Y., Anninos, P., Norman, M.L., Meiksin, A. 1996, \apj,
submitted, astro-ph 9609194
 
\refe Zheng, W., \& Davidsen,  A. 1995, \apl, 440, L53
 
\refe Zuo, L. 1992, MNRAS, 258, 36

\refe Zuo, L. \& Lu, L. 1993, \apj, 418, 601

\end{document}